\documentclass{aa}  %

\usepackage{graphicx}
\usepackage{txfonts}
\usepackage{longtable}
\begin{document}

   \title{A study of oxygen-rich post-AGB stars in the
   Milky Way to understand the production of silicates
   from evolved stars}


   \author{F. Dell'Agli\inst{1}, S. Tosi\inst{1,2}, D. Kamath$\inst{1,3,4}$, P. Ventura\inst{1,5},  H. Van Winckel\inst{6}, E. Marini\inst{2}, T. Marchetti\inst{7}
          }

   \institute{INAF, Observatory of Rome, Via Frascati 33, 00077 Monte Porzio Catone (RM), Italy \and
   Dipartimento di Matematica e Fisica, Università degli Studi Roma Tre, Via della Vasca Navale 84, 00100, Roma, Italy \and
   School of Mathematical and Physical Sciences, Macquarie University, Sydney, NSW, Australia \and
   Astronomy, Astrophysics and Astrophotonics Research Centre, Macquarie University, Sydney, NSW, Australia  \and
   Istituto Nazionale di Fisica Nucleare, section of Perugia, Via A. Pascoli snc, 06123 Perugia, Italy \and
   Institute of Astronomy, K.U.Leuven, Celestijnenlaan 200D bus 2401, B-3001 Leuven, Belgium \and 
   European Southern Observatory, Karl-Schwarzschild-Strasse 2, 85748 Garching bei München, Germany  
   }

   \date{Accepted November 21, 2022}


 \abstract
   {The study of post asymptotic giant branch (post-AGB) stars is a valuable tool to study 
   still poorly known aspects of the evolution of the stars through the asymptotic giant branch
   (AGB). This is due to the accurate determination of their surface chemical composition and
   to the peculiar shape of the spectral energy distribution (SED): the emission from the
   central star can be easily disentangled from the contribution from the dusty shell, which can
   then be characterized.}
   {The goal of the present study is to reconstruct the dust formation process and more
   generally the late phases of the evolution of oxygen-rich stars across the AGB phase. This is performed by 
   studying oxygen-rich, post-AGB stars, which are analyzed in terms of their luminosity, 
   effective temperature and infrared excess.}
   {We study sources classified as single, oxygen-rich, post-AGB stars in the Galaxy, which exhibit a 
   double-peaked (shell-type) SED. We use results from stellar evolution modelling combined with dust formation and radiative transfer modelling to reconstruct the late AGB phases and the
   initial contraction to the post-AGB phase. We also determine the mass-loss and dust formation rates
   for stars of different mass and chemical composition.}
   {The analysis of the IR excess of the post-AGB, oxygen-rich stars examined in this study 
    outlines an interesting complexity, in terms of the correlation between the dust in
    the surroundings of the stars, the evolutionary status and the progenitor's mass.
    The sources descending from massive AGBs ($>$3\,M$_{\odot}$ depending on metallicity) are generally characterized by higher infrared
    excess than the lower mass counterparts, owing to the more intense dust formation taking
    place during the final AGB phases. From the determination of the location of the dusty 
    regions we deduce that the expanding velocities of the outflow change significantly
    from star to star. The possibility that radiation pressure is unable to accelerate the 
    wind in the faintest objects is also commented.
    }
   {}


   \titlerunning{Oxygen-rich post-AGB stars in the Milky Way}
   \authorrunning{Dell'Agli et al.}
   \maketitle
%

\section{Introduction}
The stars evolving through the asymptotic giant branch (AGB) are generally regarded as
one of the most important contributors of reprocessed gas of the interstellar medium. The knowledge of 
the AGB gas yields is crucial to reconstruct the chemical evolution of the
Milky Way \citep{romano10, kobayashi20} and the chemical patterns observed in
star forming galaxies in the Local Group \citep{vincenzo16}. Furthermore, the winds of
AGB stars are a favourable site for the production of dust, which is the reason 
why the dust from AGB stars accounts for a significant fraction of the overall
dust budget in present day mature galaxies \citep{svetlana08, sloan09}. The
dust manufactured by AGB stars is a fundamental ingredient to study the dust content
of the Milky Way \citep{ginolfi18} and to model the evolution of the dust content
in local and high-redshift galaxies \citep{nanni20}.

AGB stars are extremely complex objects, where regions which can be safely
described under ideal gas conditions, such as the convective envelope, coexist with
the core, composed of carbon and oxygen (or oxygen and neon for M$\,\geq 7$\,M$_{\odot}$, \citet{garciaberro1997}), where the densities are so large that the
pressure is entirely provided by degenerate electrons \citep{busso99, herwig05, karakas14}.
Despite the significant steps forwards achieved in the last decades by different
research teams \citep{cristallo15, karakas16, karakas18, karakas22, cinquegrana, ventura18, 
ventura20, ventura21}, the description of the AGB phase is still severely affected by the
poor knowledge of physical mechanisms, mainly convection and mass loss, which play
a relevant role on the evolution of AGB stars \citep{ventura05a, ventura05b}. 

A notable contribution towards a deeper understanding of the evolution of AGB
stars is provided by the study of post-AGB stars. Indeed the surface chemical composition 
of post-AGB stars is a key indicator of the relative importance of the two mechanisms
able to alter the surface chemistry of the stars evolving through the AGB phase,
namely hot bottom burning (HBB, Bl\"ocker \& Sch\"onberner 1991) and third dredge-up 
(TDU, Iben 1974). Furthermore, owing to the  warmer photospheres (A-K spectral types) of post-AGB stars, their spectra are dominated by atomic transitions and provide the unique possibility for deriving accurate photospheric chemical abundances of a wide range of elements (e.g., CNO, alpha and Fe-peak elements, and s-process elements, see \citet{devika2020, kamath22c} and references therein). On the other hand, obtaining surface abundances for a wide range of elements using AGB star spectra is challenging since they have complex and dynamical atmospheres and their spectra are severly affected by molecular bands \citep{anibal09}. 

Typical examples of studies where observations of post-AGB are used to understand AGB nucleosynthesis are \citet{desmedt12} and \citet{devika17}. More recently, \citet{devika22a} capitalised on the reliable distance estimates derived from Gaia EDR3/DR3 to study 31 single Galactic post-AGB stars (13 oxygen-rich with no s--process enrichment and 19 carbon-rich and s--process objects) with well-studied surface abundances and derived their luminosites - a parameter that is critical to identify the initial mass of the stars. A thorough derivation of the progenitors and of the evolutionary history of these stars was presented in \citet{devika22b}.

A further advantage of the study of post-AGB stars is that the spectral energy distribution
(SED) exhibits a typical double-peak shape referred to as 'shell-type' SED \citep{hans03}, which allows to disentangle the emission from the central object and the infrared excess due to the presence of dust in the surrounding of the star. This is extremely important to test the 
dust formation modelling
in the winds of evolved stars, which was recently implemented into the description of
the AGB phase by various research teams \citep{ventura12, ventura14, nanni13, nanni14}, then
applied to interpret infrared (IR) observations and characterize AGB stars of the
Magellanic Clouds (MC) \citep{flavia14a, flavia14b, flavia15a, flavia15b, nanni16, nanni19}
and of Local Group galaxies \citep{flavia16, flavia18, flavia19, flavia21}.

The first application of this analysis has been published by \citet{tosi22} (hereafter Paper I), using the study of single post-AGB
stars to draw information regarding the previous AGB phase. This work was focused on 13 likely single sources in the MC, classified as shell-type
by \citet{devika14, devika15}, who provided the effective temperatures,
the metallicities, and a detailed reconstruction of the SED. The comparison with results 
from AGB evolution and dust formation modelling allowed to characterize the individual sources
presented in \citet{devika14, devika15} in terms of mass and formation epoch of the
progenitors, and to draw interesting conclusions on the mass loss rates experienced by
stars during the final AGB phases and on the dynamics of the outflow.

In the present work we apply the methodology proposed in Paper I to study the 13 oxygen-rich 
objects in the sample by \citet{devika22a}.
This will allow a more exhaustive investigation of the evolutionary properties of the stars
that do not become carbon stars with respect to Paper I.
Indeed, the 5 Large Magellanic Clouds (LMC) oxygen-rich stars without any s-process enrichment investigated in Paper I 
were substantially homogeneous in terms of luminosity and optical depth. The
interpretation given in Paper I was that they are the progeny of low-mass stars that
experienced no or only a few TDU episodes, which prevented any significant s-process
enrichment and the possibility to reach the carbon-star stage. 
On the other hand, the Galactic non s-process enriched sources presented in \citet{devika22a} exhibit a diversity ranging from those with large IR excesses to those without significant circumstellar dust. Moreover, compared to the LMC sample presented in Paper I, the luminosity range of the target stars in this study are much more extended (see \citet{devika22a}).

The Paper is structured as follows: the methodology followed to characterize the individual
sources examined is explained in Section\,\ref{method}; the classification of the stars in the sample
and of the dust in their surroundings is given in Section\,\ref{dust}, while Section\,\ref{disc}
is dedicated to the discussion of the dynamical properties of the outflow and on the differences
found with respect to carbon stars; finally, the conclusions are given in Section\,\ref{concl}.

\begin{figure*}
\begin{minipage}{0.32\textwidth}
\resizebox{1.\hsize}{!}{\includegraphics{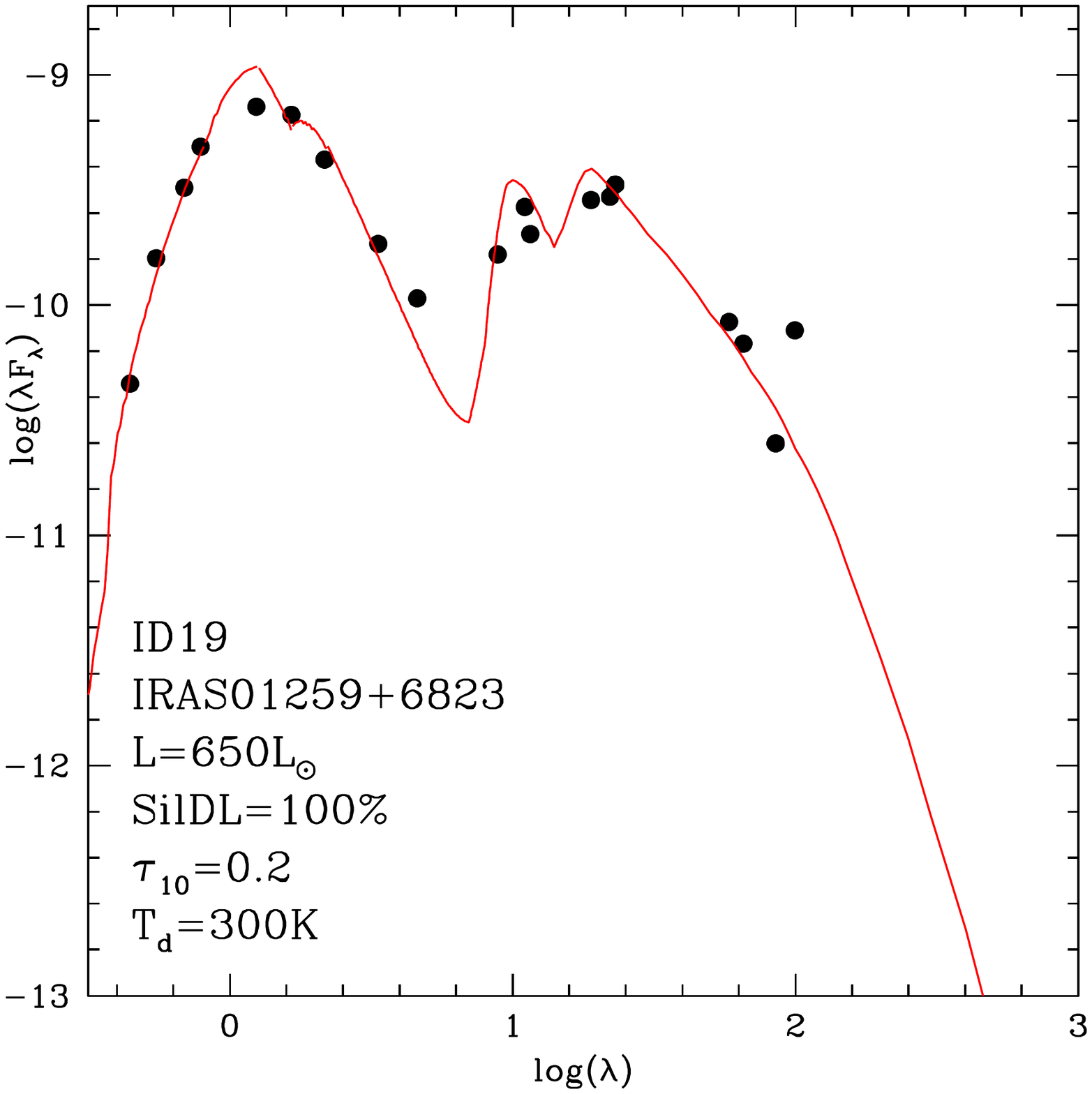}}
\end{minipage}
\begin{minipage}{0.32\textwidth}
\resizebox{1.\hsize}{!}{\includegraphics{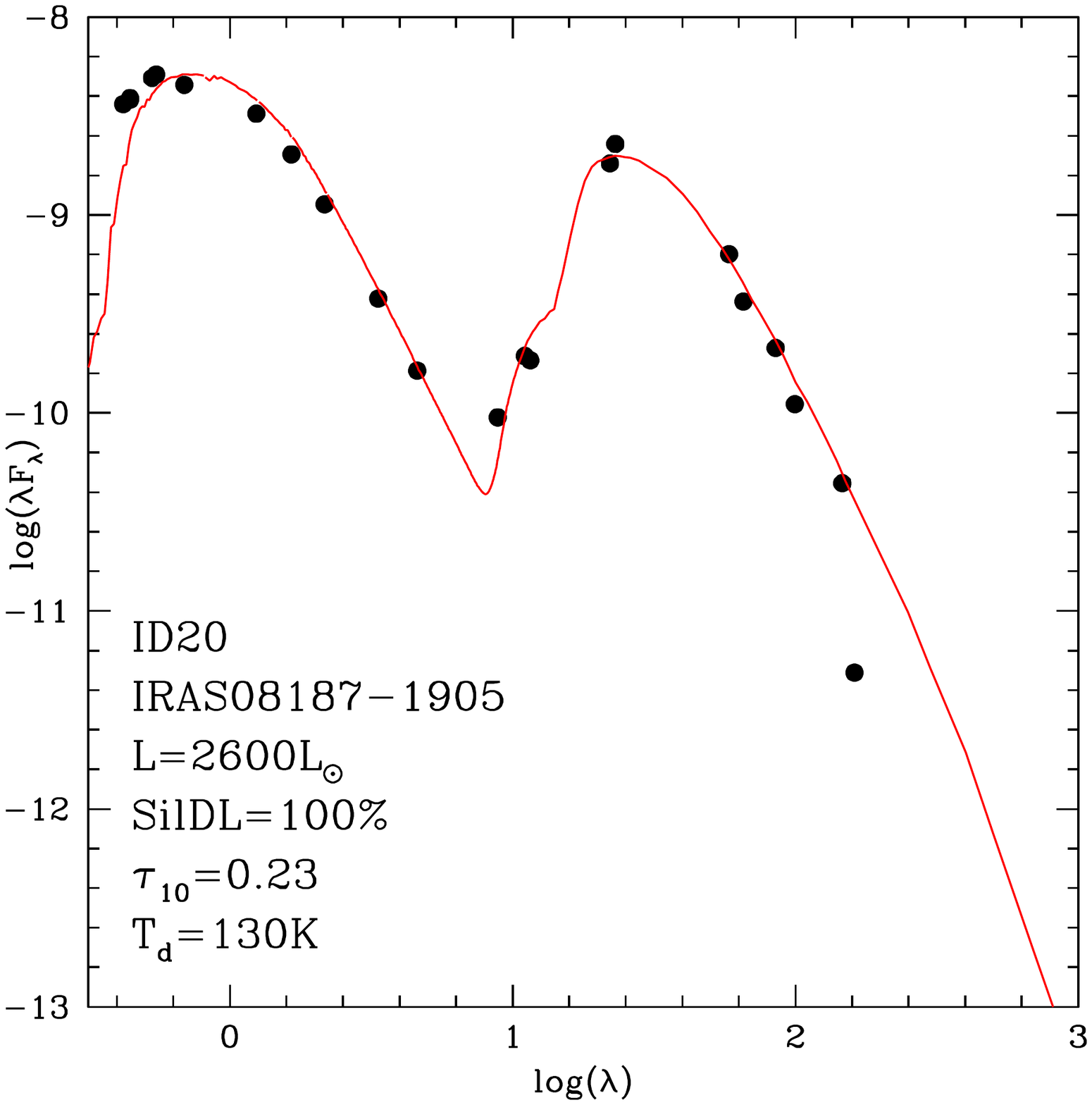}}
\end{minipage}
\begin{minipage}{0.32\textwidth}
\resizebox{1.\hsize}{!}{\includegraphics{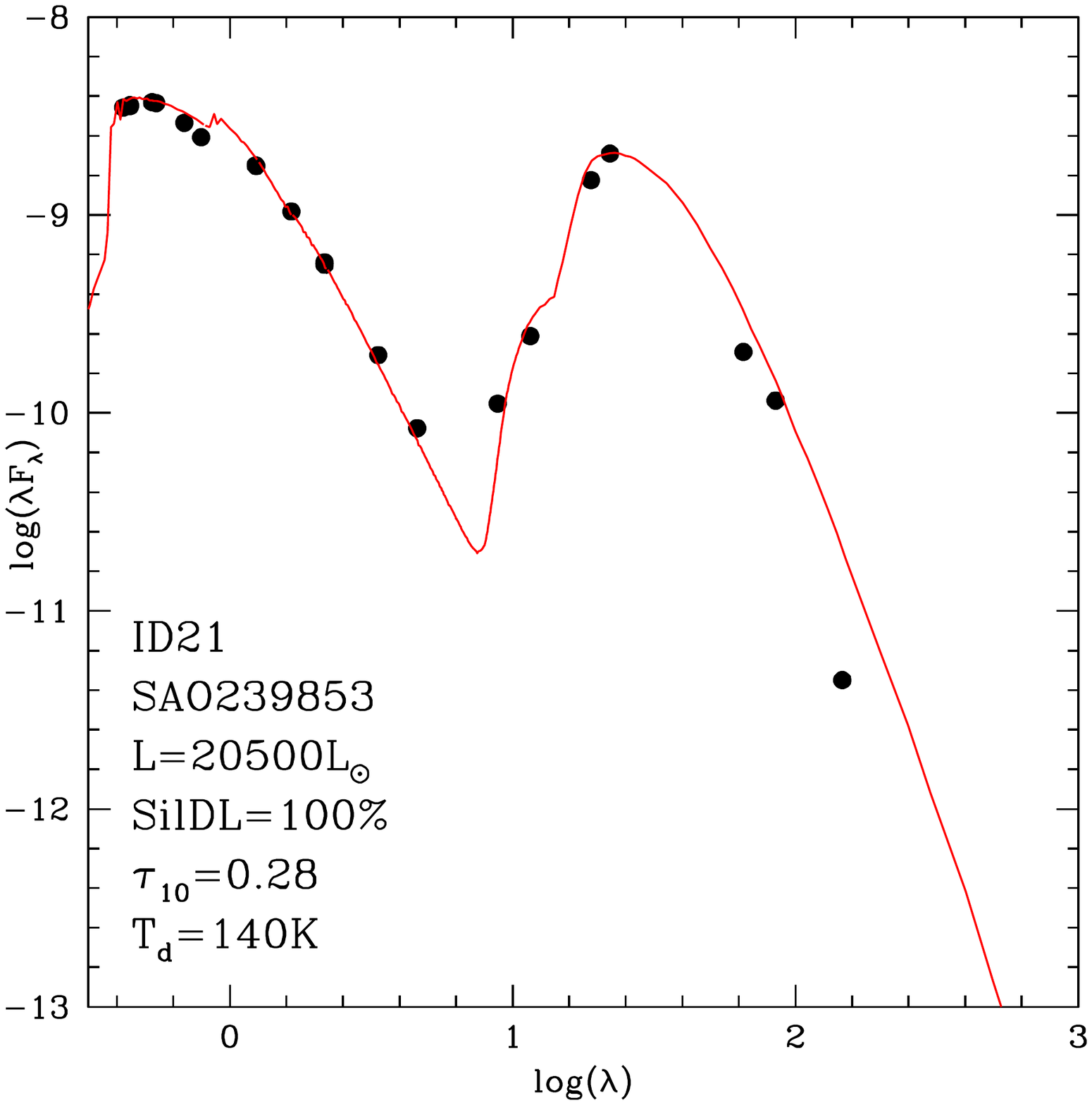}}
\end{minipage}
\vskip-70pt
\begin{minipage}{0.32\textwidth}
\resizebox{1.\hsize}{!}{\includegraphics{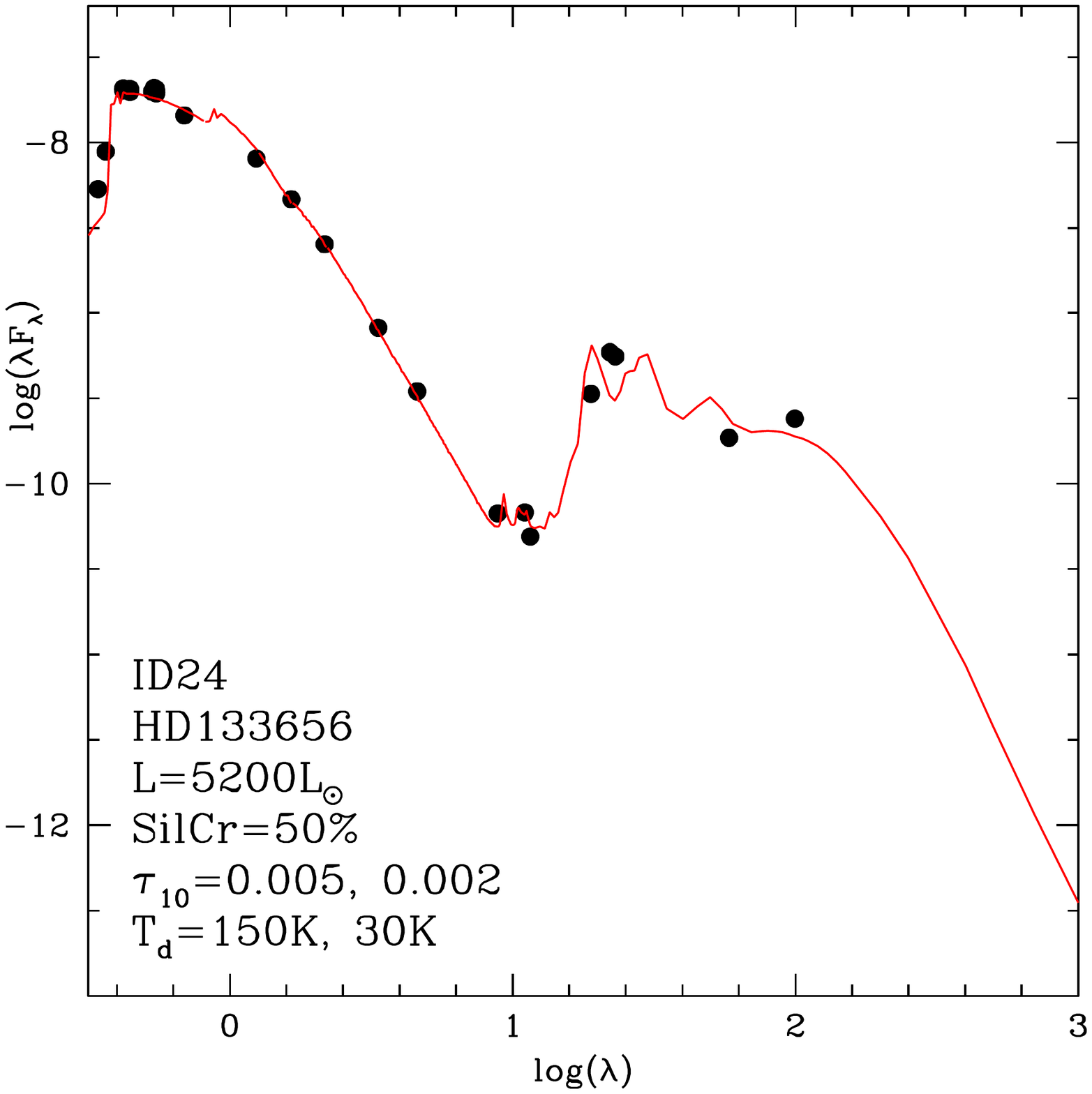}}
\end{minipage}
\begin{minipage}{0.32\textwidth}
\resizebox{1.\hsize}{!}{\includegraphics{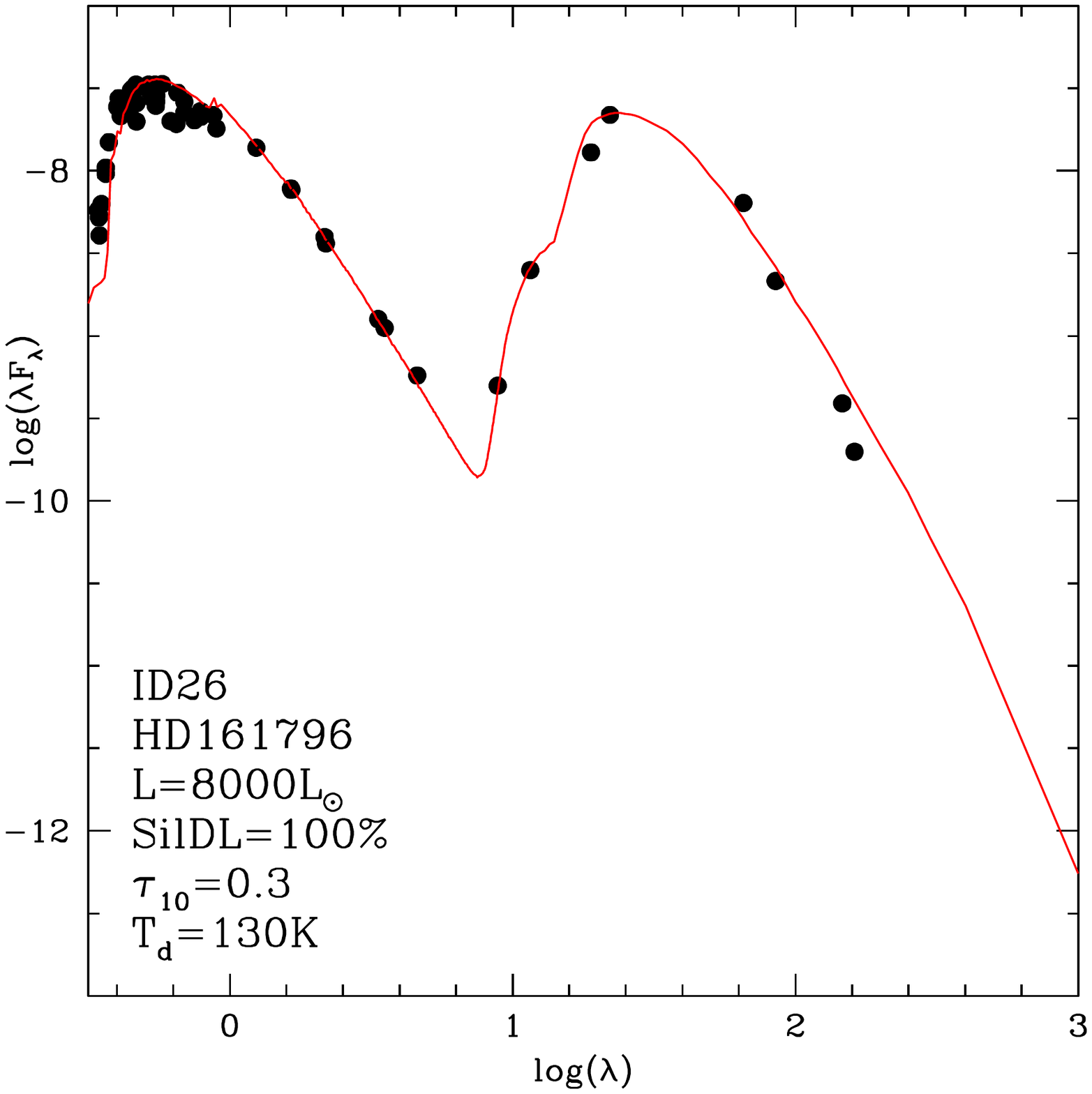}}
\end{minipage}
\begin{minipage}{0.32\textwidth}
\resizebox{1.\hsize}{!}{\includegraphics{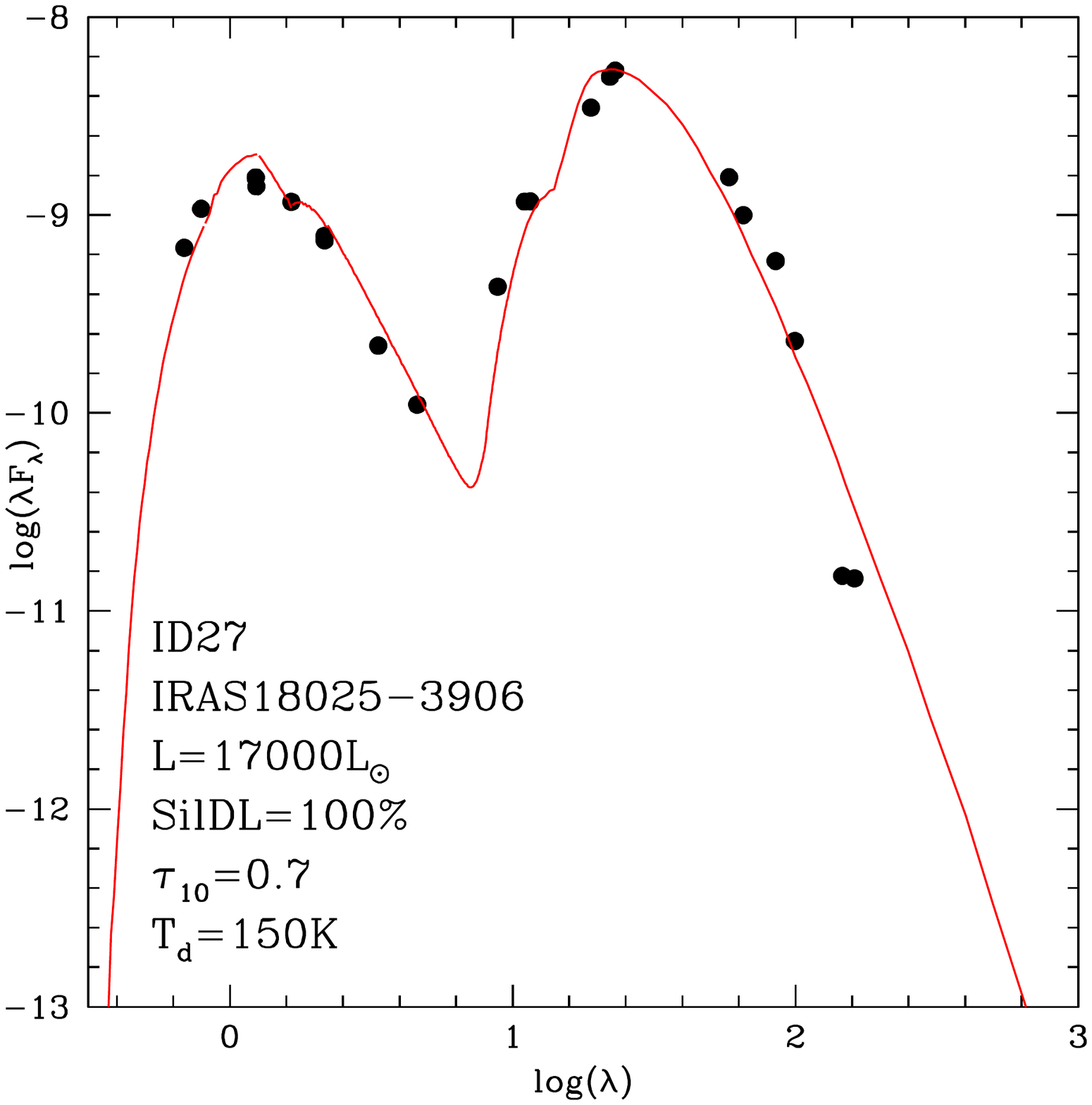}}
\end{minipage}
\vskip-70pt
\begin{minipage}{0.32\textwidth}
\resizebox{1.\hsize}{!}{\includegraphics{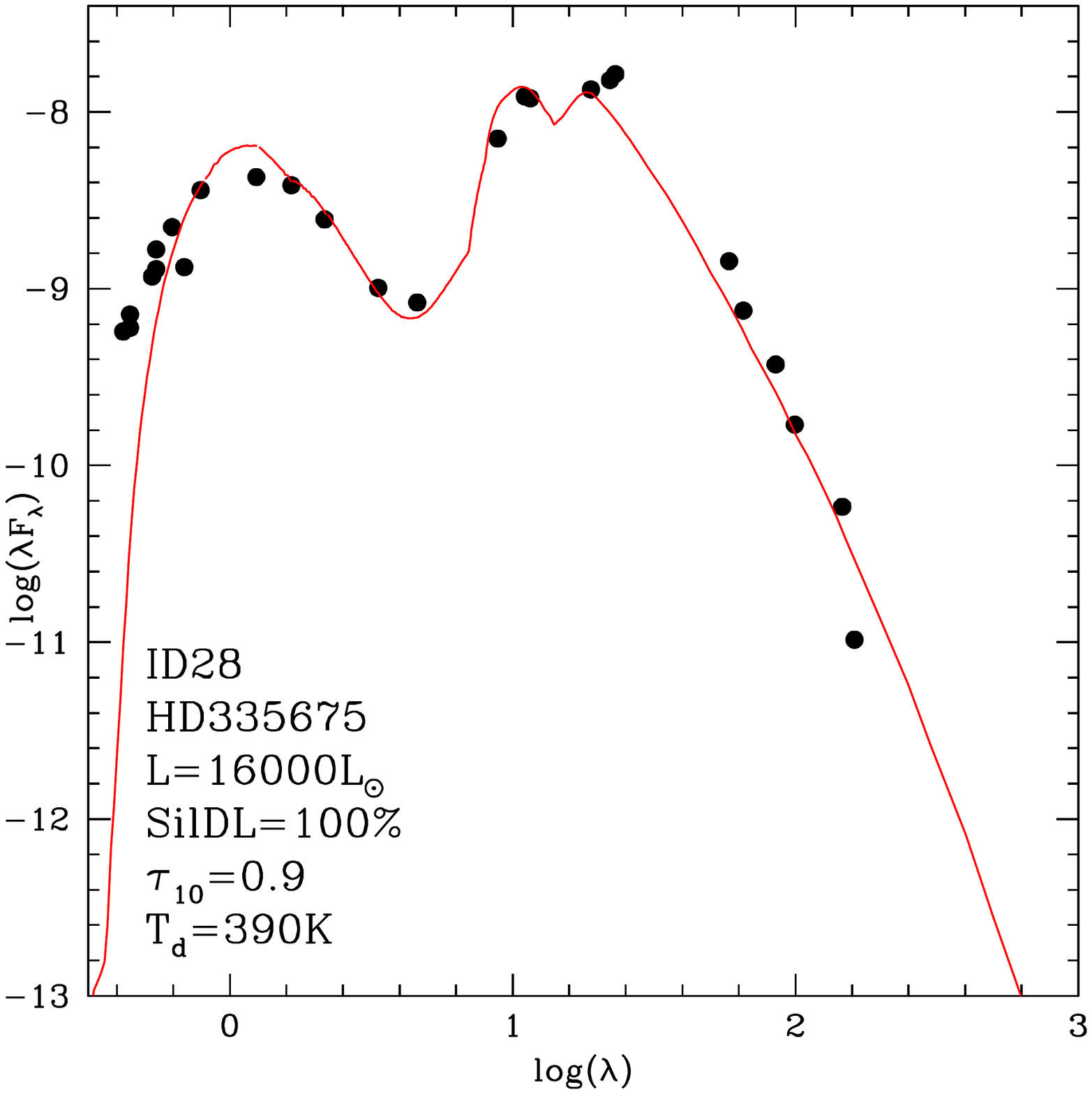}}
\end{minipage}
\begin{minipage}{0.32\textwidth}
\resizebox{1.\hsize}{!}{\includegraphics{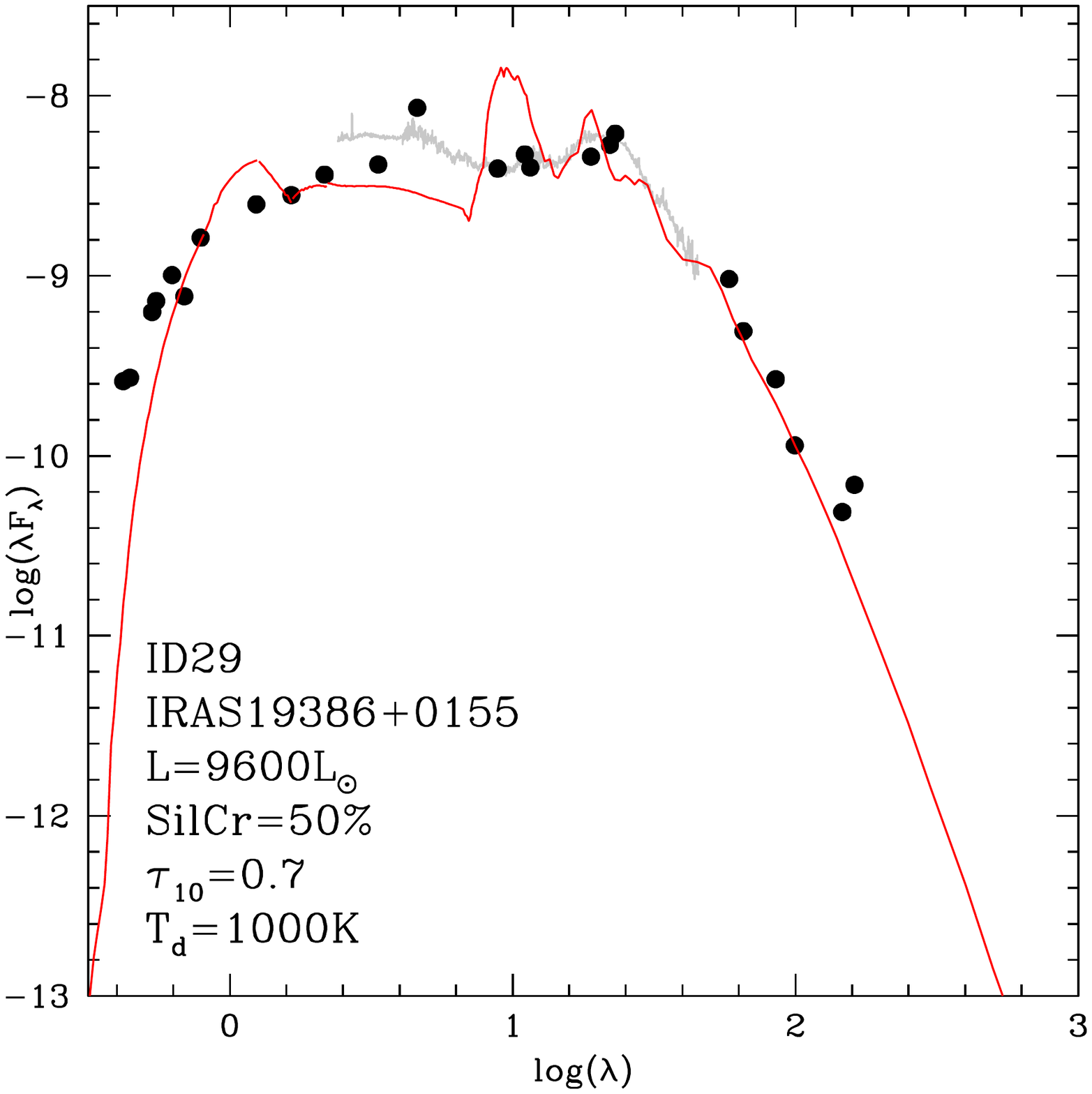}}
\end{minipage}
\begin{minipage}{0.32\textwidth}
\resizebox{1.\hsize}{!}{\includegraphics{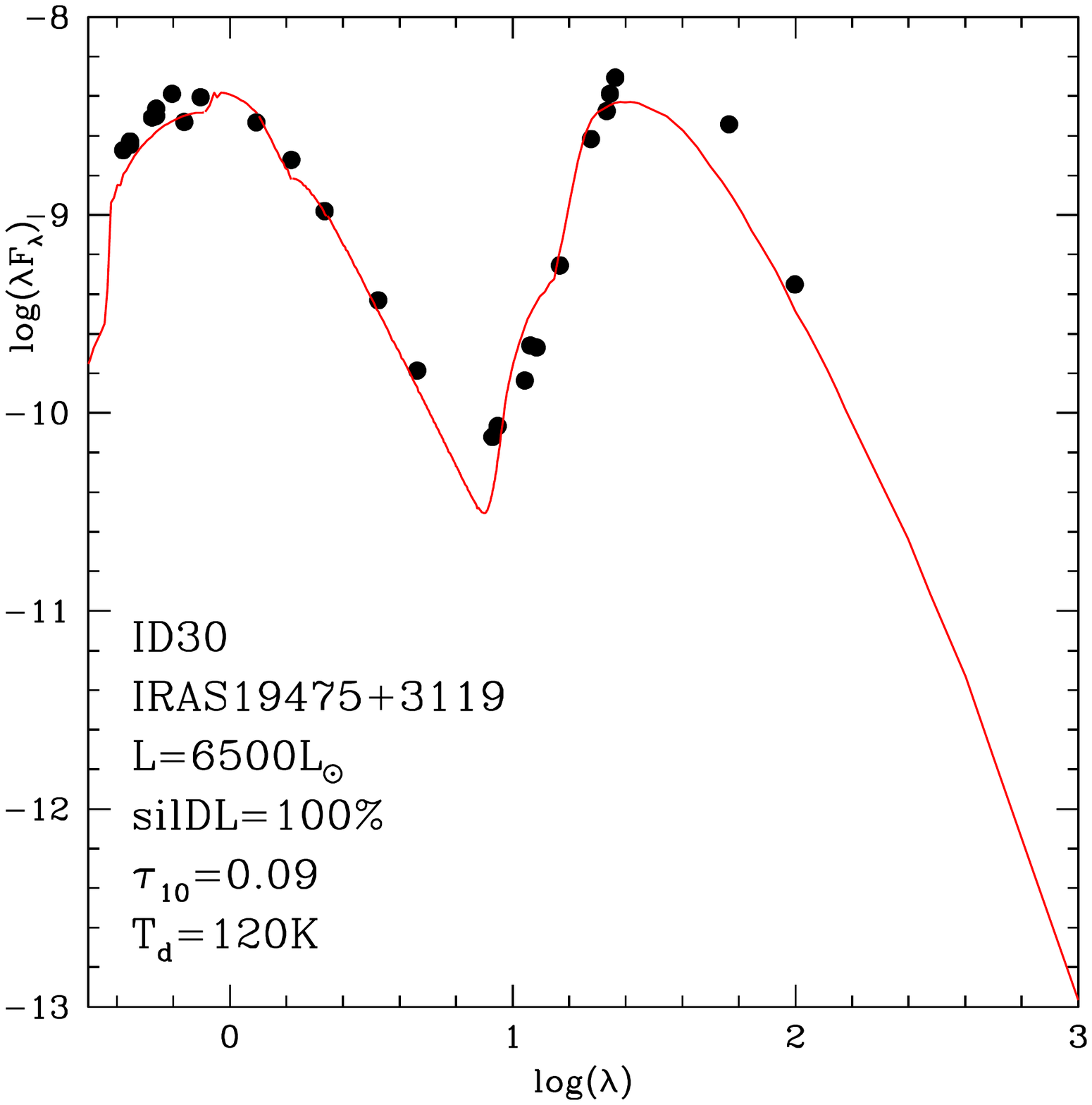}}
\end{minipage}
\vskip-40pt
\caption{Optical and IR data (black points) of oxygen-rich Galactic sources classified as post-AGB stars presented in \citet{devika22a}, which we interpret as surrounded by silicate dust in 
the present investigation. The grey line line shows
the ISO spectra from \citet{sloan03}, when available. The red lines indicate the best-fit model obtained using the DUSTY code. The derived stellar and dust parameters from this study for each source are shown in the different panels.}
\label{fsed}
\end{figure*}

\begin{figure*}
\begin{minipage}{0.24\textwidth}
\resizebox{1.\hsize}{!}{\includegraphics{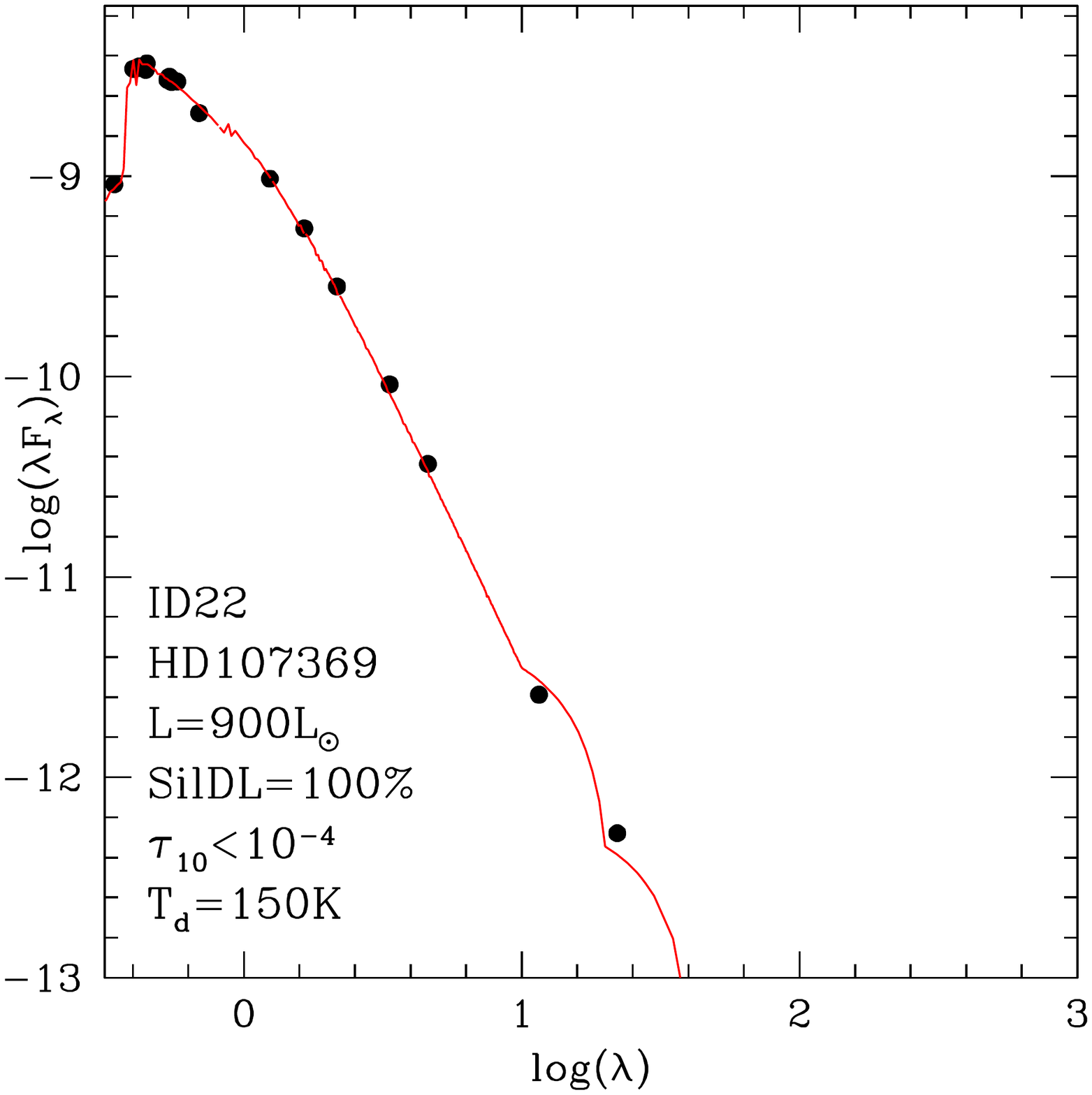}}
\end{minipage}
\begin{minipage}{0.24\textwidth}
\resizebox{1.\hsize}{!}{\includegraphics{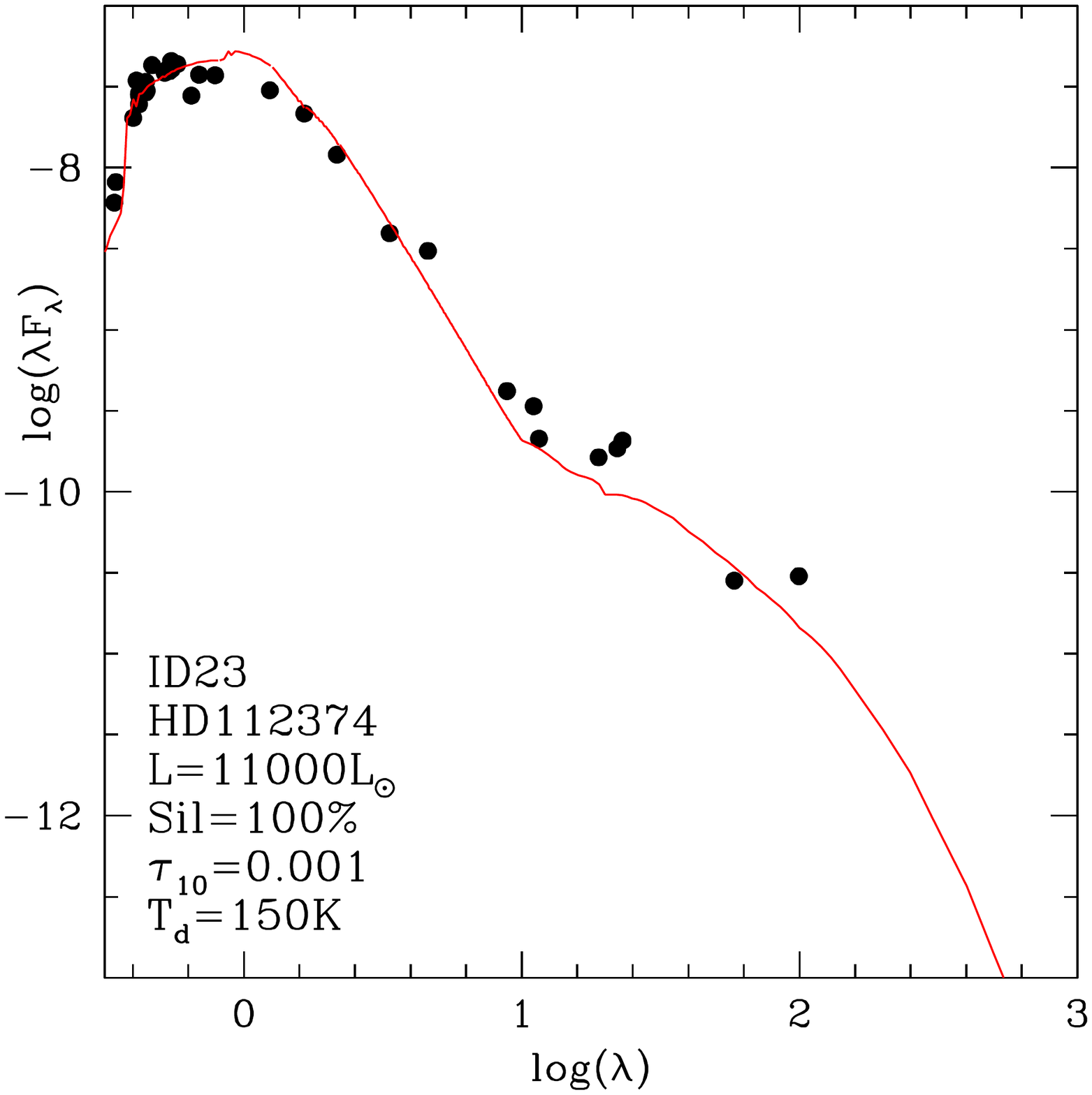}}
\end{minipage}
\begin{minipage}{0.24\textwidth}
\resizebox{1.\hsize}{!}{\includegraphics{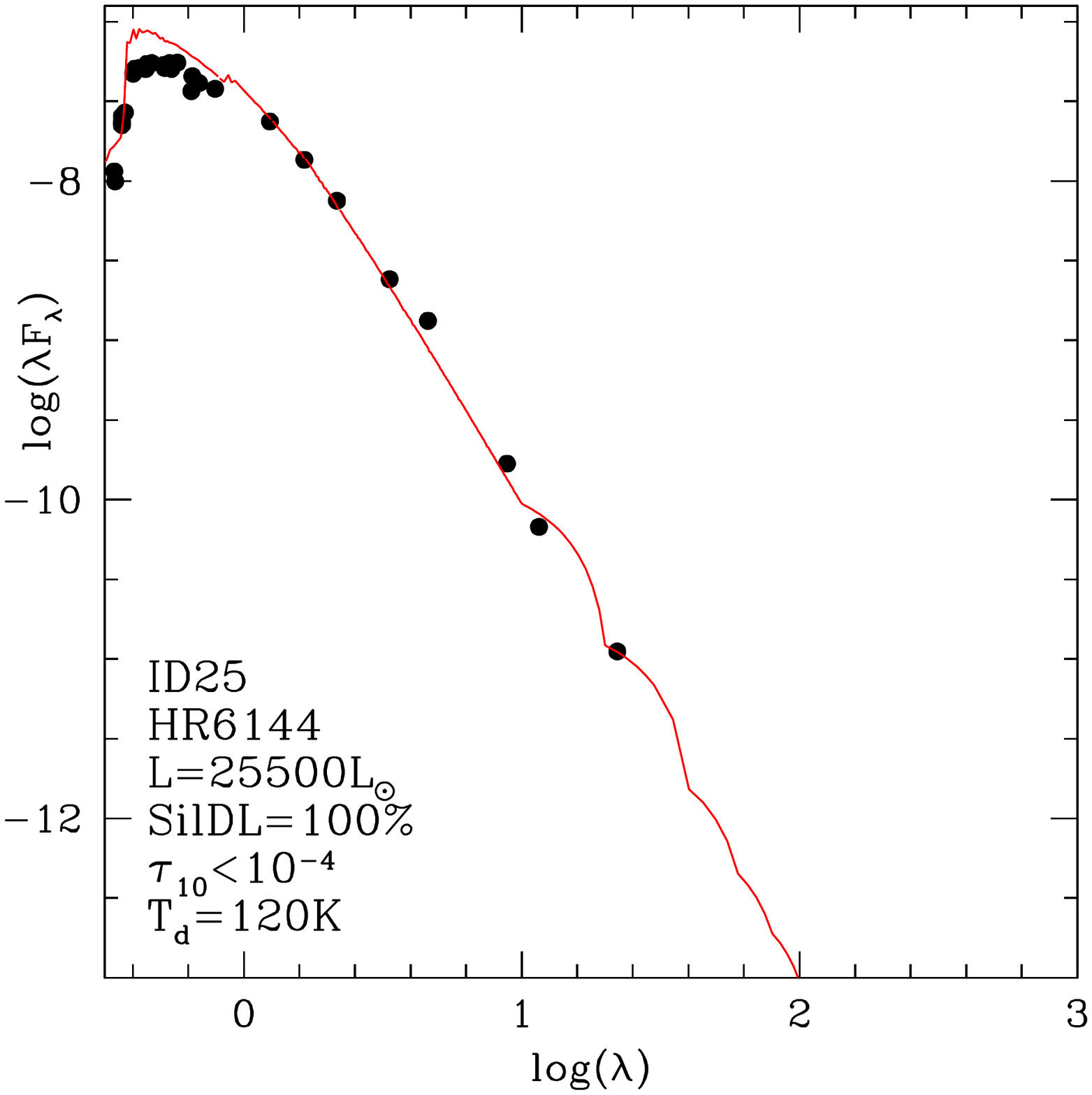}}
\end{minipage}
\begin{minipage}{0.24\textwidth}
\resizebox{1.\hsize}{!}{\includegraphics{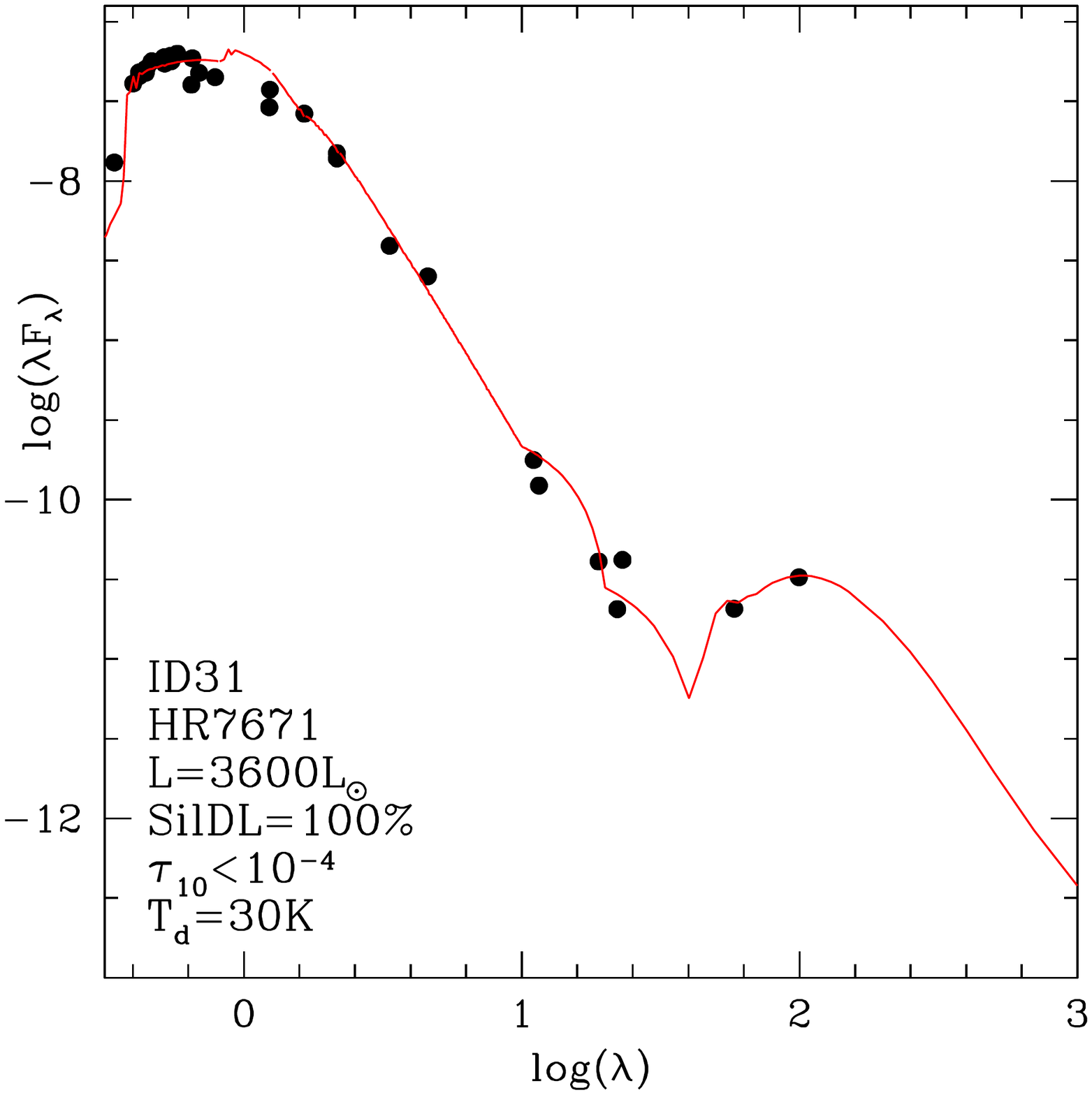}}
\end{minipage}
\vskip-20pt
\caption{Optical and IR data (black points) of oxygen-rich Galactic sources classified as post-AGB stars presented in \citet{devika22a}, which we interpret as "dust-free" in 
the present investigation. This is due to the little if not null infrared excess that characterizes these sources.}
\label{fseddustfree}
\end{figure*}

\section{The methodology to characterize the individual sources}
\label{method}
We selected the 13 oxygen-rich post-AGB stars presented by \citet{devika22a}. We applied the same
approach used in Paper I to discuss and characterize the post-AGB stars in the MC. This methodology can be summarized and divided into 3 steps:

a) The SED of each source is constructed on the basis of the photometric data collected by \citet{devika22a} and considering the \textit{Infrared Space Observatory} (ISO) spectrum \citep{sloan03}, when available. SED fitting is achieved by means of the radiative transfer code DUSTY \citep{nenkova99},
starting from the effective temperatures given in \citet{devika22a}. DUSTY is used in the spherical modality,
with the gas density declining with distance as $\propto r^{-2}$. The determination of the main physical parameters is eased by the peculiar morphology of the SED of these objects. Indeed, this procedure leads to the determination of the optical depth at 
$\lambda=10~\mu$m, $\tau_{10}$ (mainly related to the height of the peak of the SED in the mid-IR) and of ${\rm T}_d$, 
the dust temperature connected to the depth of the minimum of the SED separating the dust IR excess to the emission from the central object. Note that ${\rm T}_d$ is tightly correlated to the distance of the inner border
of the dusty zone, ${\rm R}_{\rm in}$, from the centre of the star. 
Typical uncertainties associated to these estimates are $\sim 10-15\%$ in $\tau_{10}$
and $\sim 20$ K for ${\rm T}_{\rm d}$. The distribution of the IR data provides indication on the presence of a pure amorphous silicate dust composition (optical constants from \citet{draine84}, DL hereafter) or whether a percentage of crystalline silicates (optical constants from \citet{jager94}) is present.  

b) To identify the progenitor's mass we use the evolutionary tracks of stars of different 
mass and metallicities calculated by means of the ATON code for stellar evolution \citep{ventura98}, 
thoroughly discussed in \citet{devika22b}. To this aim we compare the position of the tracks 
on the HR diagram with the effective temperatures and the luminosities of the individual sources. 
We use the metallicity and the effective temperatures given in \citet{devika22a}. Regarding the
luminosities, we note 
that unlike Paper I, focussing on MC sources, here it is required to adopt a distance for each object,
within the range given by \citet{bailer-jones2021}, which reflects into an uncertainty on the
luminosity, as found by \citet{devika22a} (see Tab. 1 in the latter paper). The choice of the distance (still within the limits mentioned above) is done by looking for consistency between the
whole observational framework of the individual stars (surface chemical abundances and dust
mineralogy) and the results from stellar evolution and dust formation modelling.
We note that the accuracy with which the distance is known sets a natural discrimination between the stars labeled as Q1 (RUWE\footnote{The RUWE is the Renormalized Unit Weight Error, which is defined as the magnitude and colour-renormalized square root of the reduced chi-squared statistic to the Gaia astrometric fit \citep{lindegren2021}. A value of RUWE < 1.4 is often employed to select stars with accurate parallaxes and proper motions, while larger values could be a sign of binarity and, in general, of a poor astrometric fit \citep[e.g. ][]{belokurov2020, penoyre2020}.}\,$<$\,1.4) in  
\citet{devika22a}, on which we will base most of our interpretation, and those flagged as Q2 (RUWE\,$\gtrsim$\,1.4), 
whose distances are much more uncertain, which reflects into a poor determination of the
luminosities.

c) To connect the properties of the dust responsible for the IR excess nowadays observed
with the dust production process during the previous AGB and early post-AGB phases,
we modelled dust formation at the tip of the AGB phase (by which we mean the stage when
the contraction to the post-AGB phase begins), and during some evolutionary stages distributed
along the post-AGB phase. This step is intended to look for consistency between the
results from stellar evolution plus dust formation modelling and the observations,
and to understand when the dust currently observed was released (see Section 5 in Paper I for 
more details). Dust formation was modelled according to \citet{ventura12, ventura14}, following 
the schematization proposed by the Heidelberg group \citep{fg06}. The input to describe dust 
formation, namely mass, effective temperature, mass loss rate and surface chemical composition 
of the star, were taken from the results of stellar evolution modelling. Because we are 
studying oxygen-rich stars only, the dust species considered are silicates (both in amorphous 
and crystalline form) and alumina dust \citep{fg06}.

\begin{figure}
\begin{minipage}{0.48\textwidth}
\resizebox{1.\hsize}{!}{\includegraphics{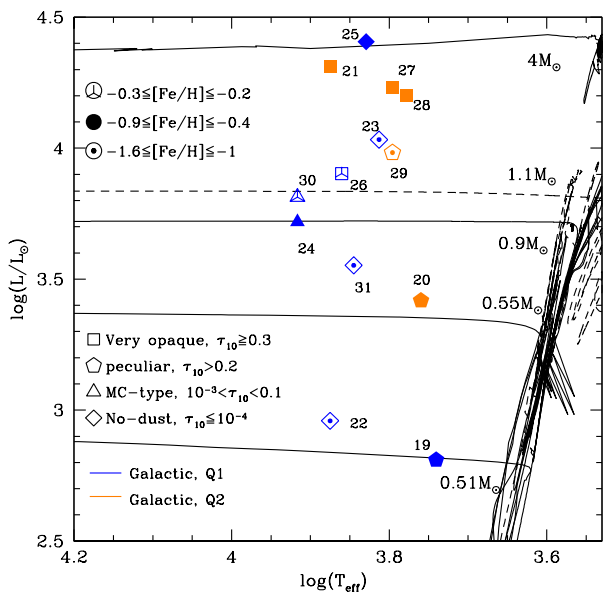}}
\end{minipage}
\vskip-60pt
\caption{Distribution of the oxygen-rich post-AGB stars studied here on the 
H-R diagram, according to the data reported in table \ref{tabpost}. 
The evolutionary tracks used for the interpretation of the individual
objects are also shown, including the late AGB and the post-AGB phase; the tracks considered have metallicity Z$=$0.004 (solid lines) and Z$=$Z$_{\odot}$ (dashed line) and their mass at the beginning of the AGB phase is reported on the right side.
}
\label{fhr}
\end{figure}

\section{The dust properties of oxygen-rich Galactic post-AGB stars}
\label{dust}

\begin{table*}
\caption{Physical and dust properties of the Galactic post-AGB stars targeted in this study.
The quantities listed in the various columns are the following:
1 - Object name; 2 - source ID (as reported in Kamath et al. 2022a); 3, 4: metallicity and effective temperatures,
derived spectroscopically by \citet{devika22a}; 5: luminosity derived in the present study for the Galactic sources and in ; 6: luminosity from \citet{devika22a} (see Table 1), obtained considering the upper and lower limits of the distances retrieved from \citet{bailer-jones2021}; 7-9: optical depth at $10~\mu$m, dust temperature and distance
separating the central star from the inner border of the dusty region, found via SED
fitting; 8: the quality flag (Q1 or Q2), based on the Gaia EDR3 renormalized unit weight error parameter, taken
from \citet{devika22a}. To ease the comparison with the results from Paper I, we report in the 5 bottom lines 
the non s-process enriched LMC sources studied in Paper I; for these sources the metallicity and the effective temperature are from \citet{devika15} while the luminosity and the dust properties are from Paper I .
}

\label{tabero}      
\centering
\begin{tabular}{l c c c c c c c c c}    
\hline      
Source & ID & $[$Fe$/$H$]$ & ${\rm T}_{\rm eff}$[K] & ${\rm L}/{\rm L}_{\odot}$ & ${\rm L}/{\rm L}_{\odot}^{K22}$ & $\tau_{10}$ & ${\rm T}_d$[K] & ${\rm R}_{\rm in}/{\rm R}_{\odot}$ & flag  \\
\hline 
\\
IRAS 01259+6823 & 19 & $-0.6 \pm 0.1$ & 5510 & 650   & 220 -- 646& 0.2   & 300  & $6.30\times 10^3$ &	Q1  \\
IRAS 08187--1905 & 20 & $-0.6 \pm 0.1$ & 5772 & 2600  & 2099 -- 3286 & 0.23  & 130  & $1.55\times 10^5$ &	Q2  \\
SAO 239853      & 21 & $-0.8 \pm 0.1$ & 7452 & 20500 & 13080 -- 48520& 0.28  & 140  & $3.80\times 10^5$ &	Q2  \\
HD 107369       & 22 & $-1.1  \pm 0.1$ & 7533 & 900  & 814 -- 1010&   $<10^{-4}$   &  150   & -- &	Q1  \\
HD 112374       & 23 & $-1.2  \pm 0.1$ & 6393 & 11000 & 9961 -- 11882& $\leq10^{-3}$ &  150   & $6.45\times 10^4$ &	Q1  \\
HD 133656       & 24 & $-0.7 \pm 0.1$ & 8238 & 5200  & 4861 -- 5690& 0.005 & 150  & $2.00\times 10^5$ &	Q1  \\
HR 6144         & 25 & $-0.4 \pm 0.1$ & 6728 & 25500 & 22212 -- 30419 &  $<10^{-4}$  &  120   & -- &	Q1  \\
HD 161796       & 26 & $-0.3 \pm 0.1$ & 6139 & 8000  & 5209 -- 6322 & 0.3   & 130  & $2.85\times 10^5$ &	Q1  \\
IRAS 18025--3906 & 27 & $-0.5 \pm 0.15$& 6154 & 17000  & 975 -- 12963 & 0.7   & 150  & $3.60\times 10^5$ &  Q2  \\
HD 335675       & 28 & $-0.9 \pm 0.2$ & 6082 & 16000 & 7303 -- 28359 & 0.9   & 390  & $3.95\times 10^4$ &	Q2  \\
IRAS 19386+0155 & 29 & $-1.1  \pm 0.14$& 6303 & 9600  & 4345 -- 22765 & 0.7   & 1000 & $6.00\times 10^3$ &	Q2  \\
IRAS 19475+3119 & 30 & $-0.24 \pm 0.15$& 8216 & 6500  & 5955 -- 7545 & 0.09  & 120  & $4.15\times 10^5$ &	Q1  \\
HR 7671         & 31 & $-1.6  \pm 0.1$ & 6985 & 3600  & 3449 -- 3734 &  $<10^{-4}$  &  30  & -- &	Q1  \\
\\
\hline
\\
J052740.75--702842.0 & LMC & -0.50 & 8283 &	5200  & -- &	0.024 & 470 &	$1.22\times 10^4$ &	-- \\
J045119.94--670604.8 & LMC & -0.40 & 8280 &	5000  & -- &	0.006 & 270 & $2.97\times 10^4$ &	-- \\
J052241.52--675750.2 & LMC & -0.50 & 8284 &	5000  & -- &	0.017 & 400 &	$2.26\times 10^4$ &	-- \\ 
J050221.17--691317.2 & LMC & -0.60 & 5250 &	4500  & -- &	0.022 & 180 &	$5.45\times 10^4$ &	-- \\ 
J051906.86--694153.9 & LMC & -1.30 & 5613 &	3800  & -- &	0.015 & 180 &	$5.13\times 10^4$ &	-- \\
\\
\hline
\label{tabpost}
\end{tabular}
\end{table*}

The fit of the SED of the different sources considered here is shown in Fig.~\ref{fsed} 
and Fig.~\ref{fseddustfree}. A summary of the results obtained is reported in 
Table \ref{tabpost}, which also gives the metallicity, the effective temperatures 
and the luminosity range of each source, derived by \citet{devika22a}. 
The IDs in col.~2 are the same as in \citet{devika22a} and \citet{devika22b}.
The distribution of the star in the HR diagram is shown in Fig.~\ref{fhr},
where also the evolutionary tracks of some model stars used later in this section
to characterize the sample sources are also indicated.

\begin{figure*}
\begin{minipage}{0.48\textwidth}
\resizebox{1.\hsize}{!}{\includegraphics{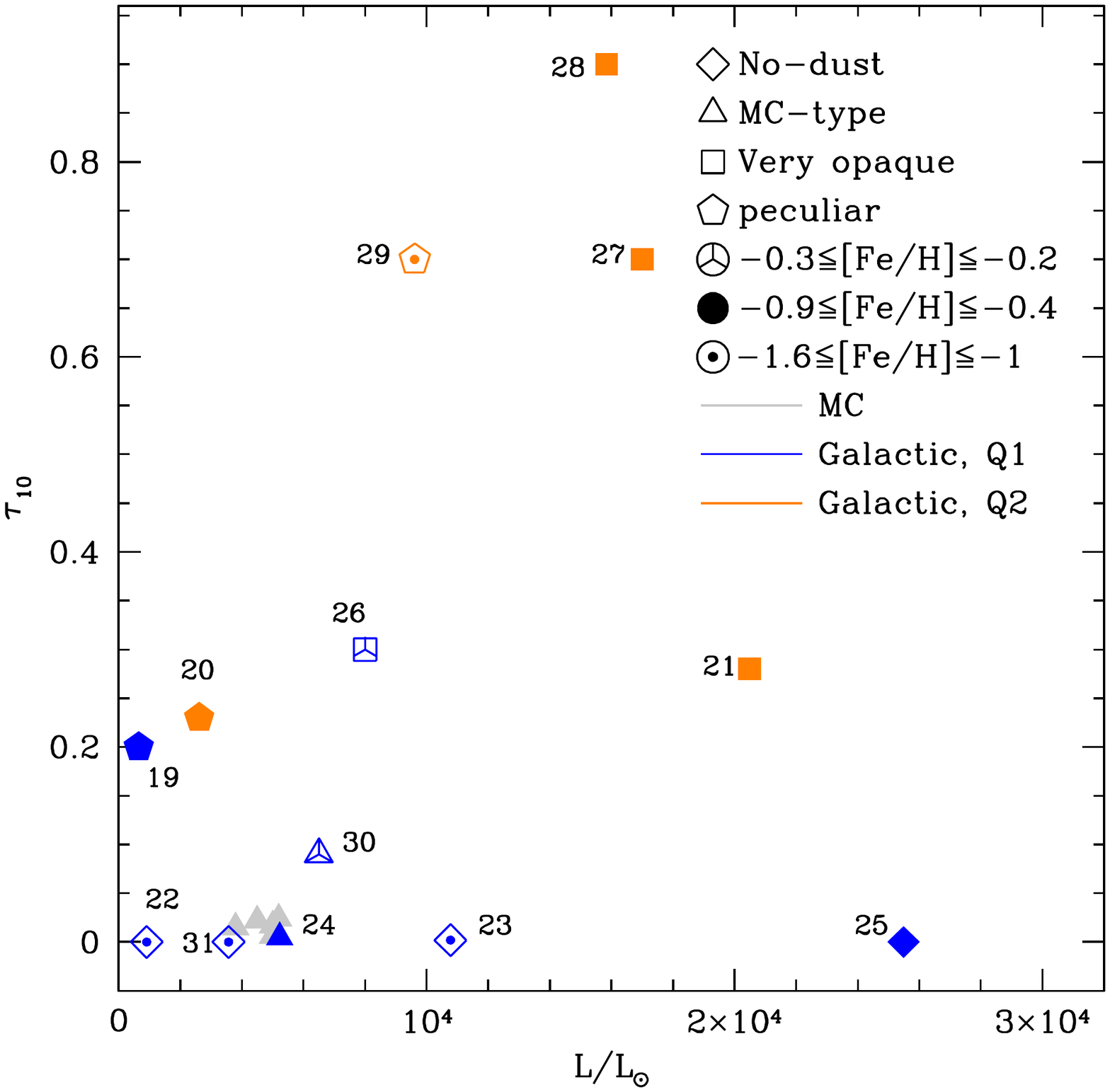}}
\end{minipage}
\begin{minipage}{0.48\textwidth}
\resizebox{1.\hsize}{!}{\includegraphics{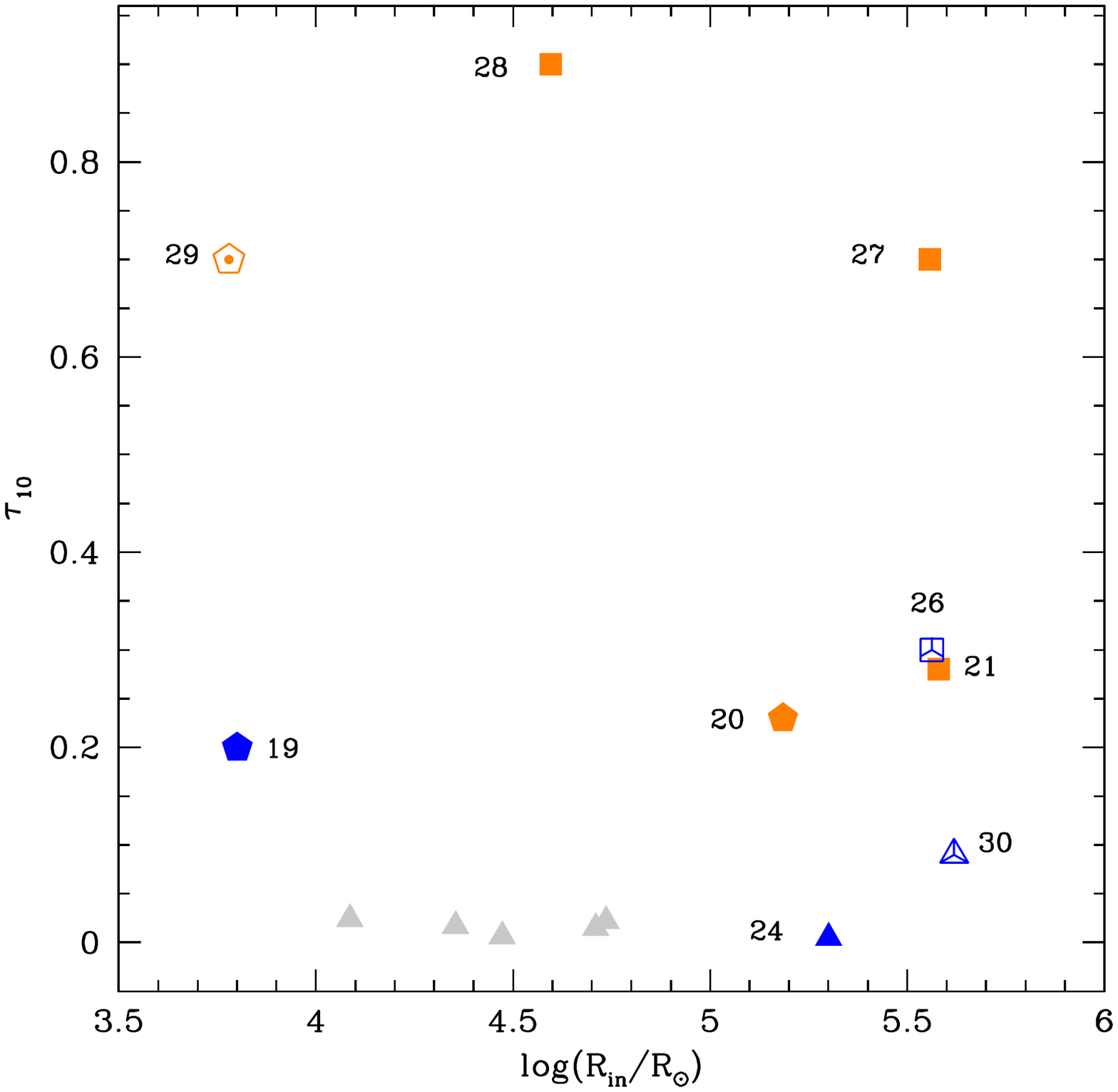}}
\end{minipage}
\vskip-60pt
\caption{The optical depths at the wavelength $\lambda = 10~\mu$m derived from SED fitting in Section
\ref{dust}, as a function of the luminosity of the star (left panel) and of the distance of the
inner border of the dusty zone from the centre of the star (right panel). Colour coding distinguishes
the Galactic post-AGB stars flagged as Q1 (blue points) from those flagged as Q2 (orange). 
The grey triangles in the lower part of the plane refer to the LMC oxygen-rich post-AGB stars studied in Paper I.}
\label{fsumm}
\end{figure*}

Similarly to Paper I, to discuss the results obtained we analyze 
the trend of $\tau_{10}$ as a function of the luminosity and of
${\rm R}_{\rm in}$, shown, respectively, in the left and right panels of Fig.~\ref{fsumm}. In the following part of 
this Section we will consider groups of stars sharing similar dust properties 
or progenitor's mass, and discuss them separately.

\begin{figure*}
\begin{minipage}{0.48\textwidth}
\resizebox{1.\hsize}{!}{\includegraphics{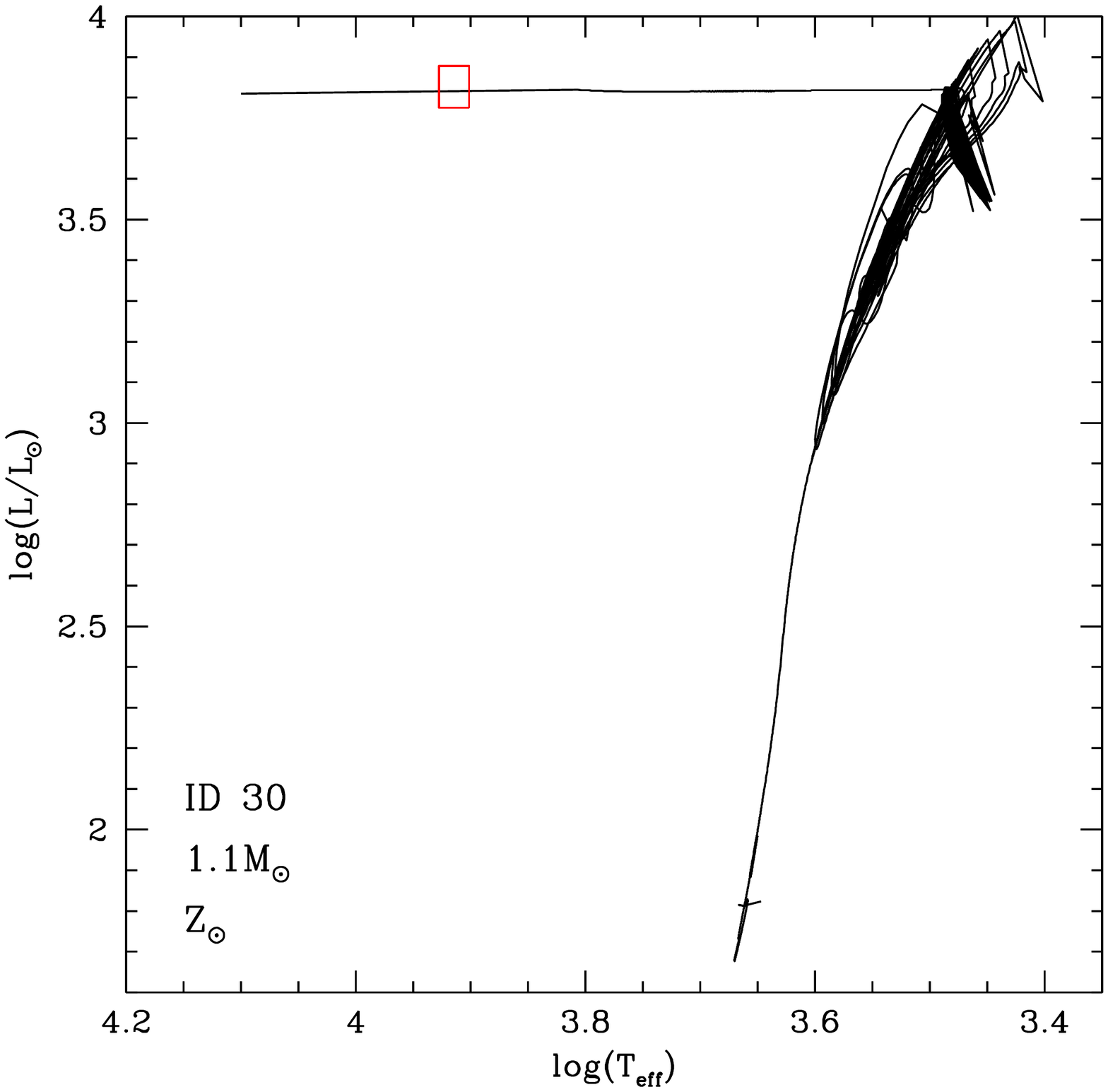}}
\end{minipage}
\begin{minipage}{0.48\textwidth}
\resizebox{1.\hsize}{!}{\includegraphics{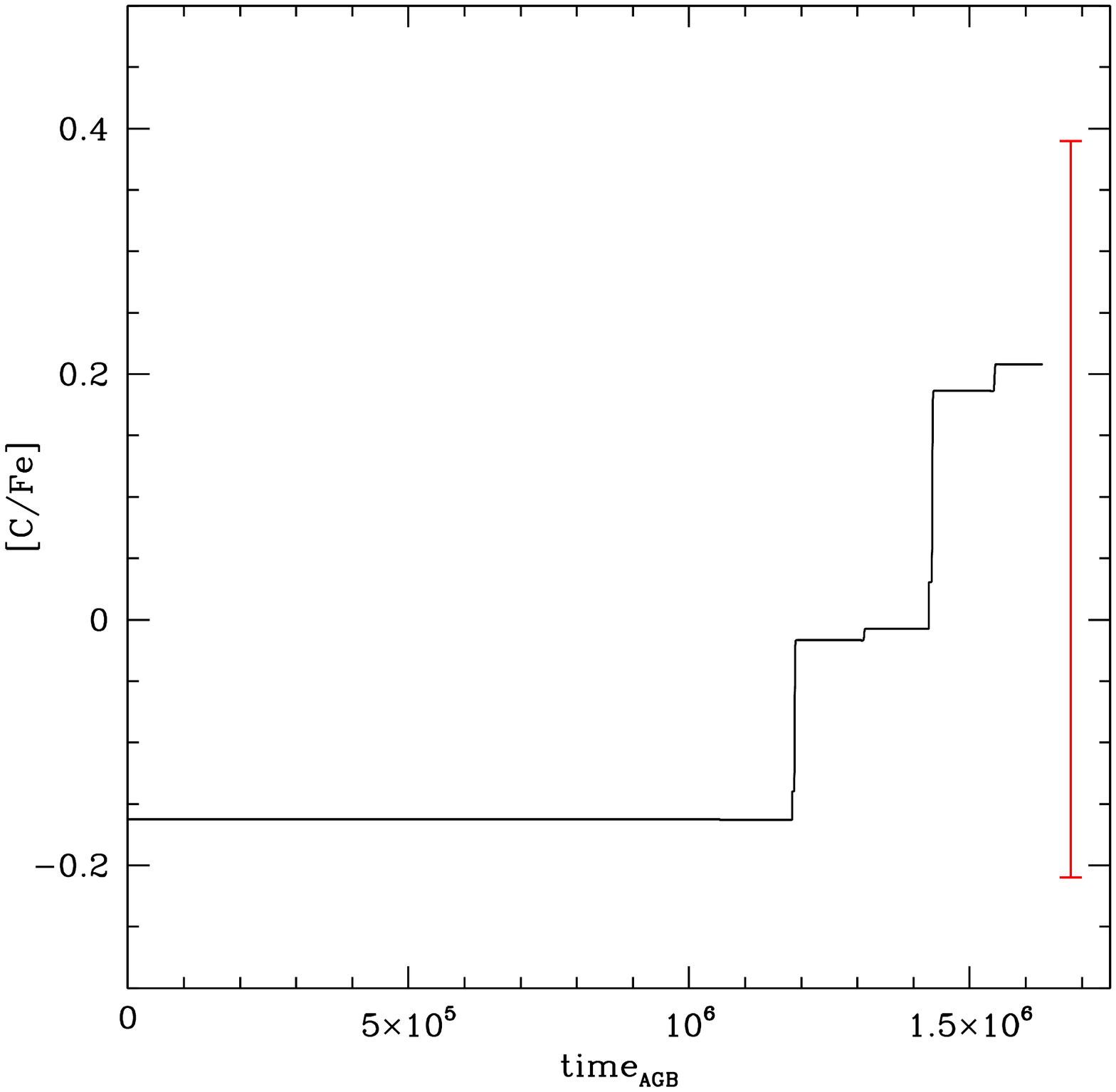}}
\end{minipage}
\vskip-60pt
\caption{The excursion of the evolutionary track of a $1.1~{\rm M}_{\odot}$
star of solar metallicity (left panel) and the variation of the surface
carbon of the same $1.1~{\rm M}_{\odot}$ model star during the AGB phase.
The red box on the left panel indicates the effective temperature and
luminosity (with the corresponding error bars) given for ID 30 in \citet{devika22a},
whereas the red vertical line in the right panel indicates the 
surface $[$C$/$Fe] of ID 30. 
}
\label{fid30}
\end{figure*}

\subsection{The oxygen-rich$/$carbon star interface}
The stars ID 24 and ID 30 were interpreted by \citet{devika22b} as low-mass
stars that lose the envelope after only a few TDU events, and did not turn
into carbon stars (see Fig.~2 in \citet{devika22b}). Consistently
with this interpretation, in the left panel of Fig.~\ref{fsumm} the two sources
populate the same region as the 4 LMC oxygen-rich stars discussed in Paper I and for this reason we will refer to them as "MC-type" stars. In Paper I they were also identified as low-mass stars that did not reach the C-star stage 

In this group ID 30 is the source with the highest $\tau_{10}$ ($\tau_{10}=0.09$, while
$\tau_{10} \sim 0.03$ on the average), which is the effect of the higher luminosity and 
the higher mass-loss rate experienced at the end of the AGB phase. The luminosity range of ID 30 given 
in \citet{devika22a}, namely $5955-7545~{\rm L}_{\odot}$, is consistent with the 
interpretation given in \citet{devika22b} only if values close to the lower limit are 
considered (see Fig.~3 in \citet{devika22b}).

To test whether a satisfactory agreement between the physical and dust parameters
derived from the observations and from SED fitting with the results from
stellar evolution + dust formation modelling can be reached, we investigated
ID 30 in detail. Because the metallicity of this star is higher than the other low-luminosity 
sources for which we propose a similar origin, all sharing a sub-solar chemical composition,
we considered solar metallicity models, specifically calculated for the present investigation. 

The left panel of Fig.~\ref{fid30} shows the evolutionary track of a $1.1~{\rm M}_{\odot}$
model star of solar metallicity. Note that in agreement with \citet{devika22b} the value
of the mass refers to the beginning of the core helium burning phase, from which the
evolutionary run is started. The post-AGB luminosity of this model-star is 
$\sim 6700~{\rm L}_{\odot}$, which is within the luminosity range given by
\citet{devika22a}. Overall, this star experiences 10 thermal pulses (TPs) and only 2 TDU events.
As shown in the right panel of Fig.~\ref{fid30}, the surface carbon increases
up to $[$C$/$Fe$]\sim 0.2$, consistently with the results given in \citet{devika22a},
far below the threshold required to convert the star into a carbon star. 

If we assume that the star lost $0.1~{\rm M}_{\odot}$ during the RGB, we derive
a progenitor initial mass of $1.2~{\rm M}_{\odot}$, which corresponds to an age for
ID 30 of $\sim 6$ Gyr. The uncertainty associated to the derivation of the post-RGB mass of ID 30 
given above is small. Indeed if we consider masses $\sim 1~{\rm M}_{\odot}$ we find 
post-AGB luminosities smaller than the lower limit for ID 30 given in \citet{devika22a};
on the contrary, if we assume a higher mass we find that the star experiences
a higher number of TDU events than those shown in the right panel of Fig.~\ref{fid30},
and eventually becomes a carbon star, with significant s-process enrichment.

Fig.~\ref{fid30} shows that the $1.1~{\rm M}_{\odot}$ model star reaches a 
maximum size slightly in excess of $300~{\rm R}_{\odot}$ before starting the
contraction to the post-AGB phase. The mass loss rate during the phase when the radius
is largest, to which we refer to as tip of AGB (TAGB), is $\sim 2\times 10^{-6}~{\rm M}_{\odot}/$yr.
This is $\sim 2$ times higher than the mass-loss rates of the low-luminosity oxygen-rich 
stars studied in Paper I, which is due to the higher luminosity and the larger metallicity. 

Our dust formation modelling applied to the TAGB parameters found for ID 30
(${\rm L} = 6700~{\rm L}_{\odot}$, $\dot {\rm M} = 2\times 10^{-6}~{\rm M}_{\odot}$/yr,
effective temperature 3000K) leads to a nowadays optical depth 
of $\tau_{10} = 0.5$, significantly higher than the value derived from SED fitting.
The possibility that the dust was released at the tip of the AGB can be also
disregarded on the basis of dynamical arguments: indeed the time required for the
star to evolved from the TAGB to the current phase is found to be $\sim 10^4$ yr,
which, combined with the distance of the dusty region of ID 30 reported in 
table \ref{tabpost}, would imply velocities below 1 km$/$s, inconsistent with 
results from the observations and from dynamical modelling of radiation-driven
stellar winds.
On the other hand, if we assume that the dust responsible for the IR excess 
observed now was released when the effective temperature of the star reaches
$3500$ K, we find that $\tau_{10} \sim 0.1$, in agreement with the results
shown in Fig.~\ref{fsed}, reported in Table \ref{tabpost}. 
Considering that the time between the phase when the effective temperature
was 3500 K and the present epoch is found to be slightly below 1 Kyr,
by assuming an expansion velocity of 10 km$/$s, we find that the dusty layer 
should currently be at a distance $\sim 4\times 10^5~{\rm R}_{\odot}$ away from 
the central star, consistent with the analysis from SED fitting 
(see right panel of Fig.~\ref{fsumm}).

When compared to the 5 low-luminosity, oxygen-rich stars investigated in Paper I 
(grey triangles in Fig.~\ref{fsumm}), it is clear that the dust has travelled 
further in the case of ID 30, reaching distances $3-10$ times higher. We propose that 
this is related to the different optical depths, which witnesses a significant larger 
dust formation in the case of ID 30. In the stars discussed in Paper I the amount of 
dust manufactured was so small that the effects of the radiation 
pressure are reduced, thus the velocities with which the outflow moved away from the star were
of the order of a few km$/$s. In the case of ID 30 dust was produced in higher
quantities, which favoured a stronger action from radiation pressure, which in turn
reflects into higher velocities of the outflow. Considering that ID 30 has higher luminosity and metallicity with respect to the LMC counterparts, this interpretation is consistent with the study by \citet{goldman2017}, who found that the expansion velocity of the outflow of oxygen-rich stars in the LMC and in the Galaxy are positively correlated with the metallicity and luminosity. 

The source ID 24 is characterized by lower luminosity and metallicity than ID 30.
The post-AGB luminosity of the star is reproduced by a $0.9~{\rm M}_{\odot}$ star
of metallicity $Z=4\times 10^{-3}$. By accounting for the fact that $0.1-0.2~{\rm M}_{\odot}$
were lost during the RGB phase, we deduce that this star formed 5-6 Gyr ago.
The contraction time from the tip of the AGB to the current effective temperature
${\rm T}_{\rm eff}=8238$ K is $2.5\times 10^4$ yr, whereas the time from the stage
when ${\rm T}_{\rm eff}=3500$ K to nowadays is 2500 Kyr. This time scale is consistent
with the distance of the dusty region derived from SED fitting, if we assume that the
dusty layer travelled with a velocity $\sim 2$ Km$/$s.

\subsection{Post-AGB stars that experienced HBB}
\citet{devika22b} identified the sources ID21 and ID 27 as the progeny of massive AGB stars that experienced HBB. This conclusion is based on the surface chemical composition, which shows up signs of nitrogen enrichment, whereas the [C/Fe] is slightly super solar. The analysis of these sources offers the possibility to 
investigate dust production by massive AGBs, something which was not possible 
in the study of Paper I. This approach proves useful also to constrain the mass loss suffered during the late AGB phases by the stars which experience HBB. Unfortunately they are both flagged as Q2, thus their distance is highly
uncertain, so their luminosity. As far as ID 21 is concerned, the post-HBB 
interpretation is consistent with the luminosity range given in \citet{devika22a}; 
for what concerns ID 27, the luminosity range given by \citet{devika22a} is
$\sim 1000-13000~{\rm L}_{\odot}$, while the luminosities of post-HBB stars are
expected to be above $\sim 17000~{\rm L}_{\odot}$; however, given the Q2 flag, 
we believe that the luminosity of this source is underestimated.

\begin{figure*}
\begin{minipage}{0.48\textwidth}
\resizebox{1.\hsize}{!}{\includegraphics{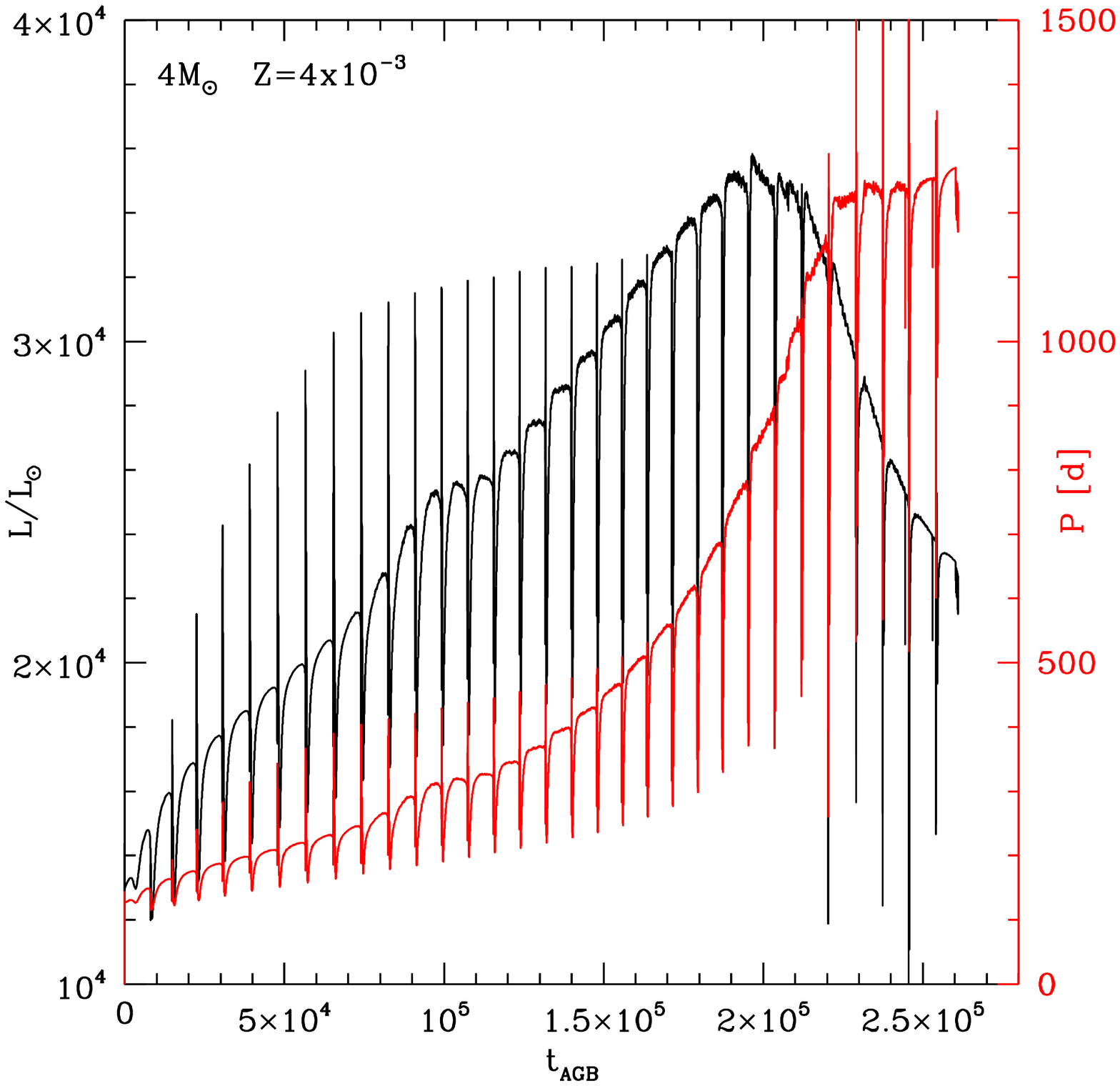}}
\end{minipage}
\begin{minipage}{0.48\textwidth}
\resizebox{1.\hsize}{!}{\includegraphics{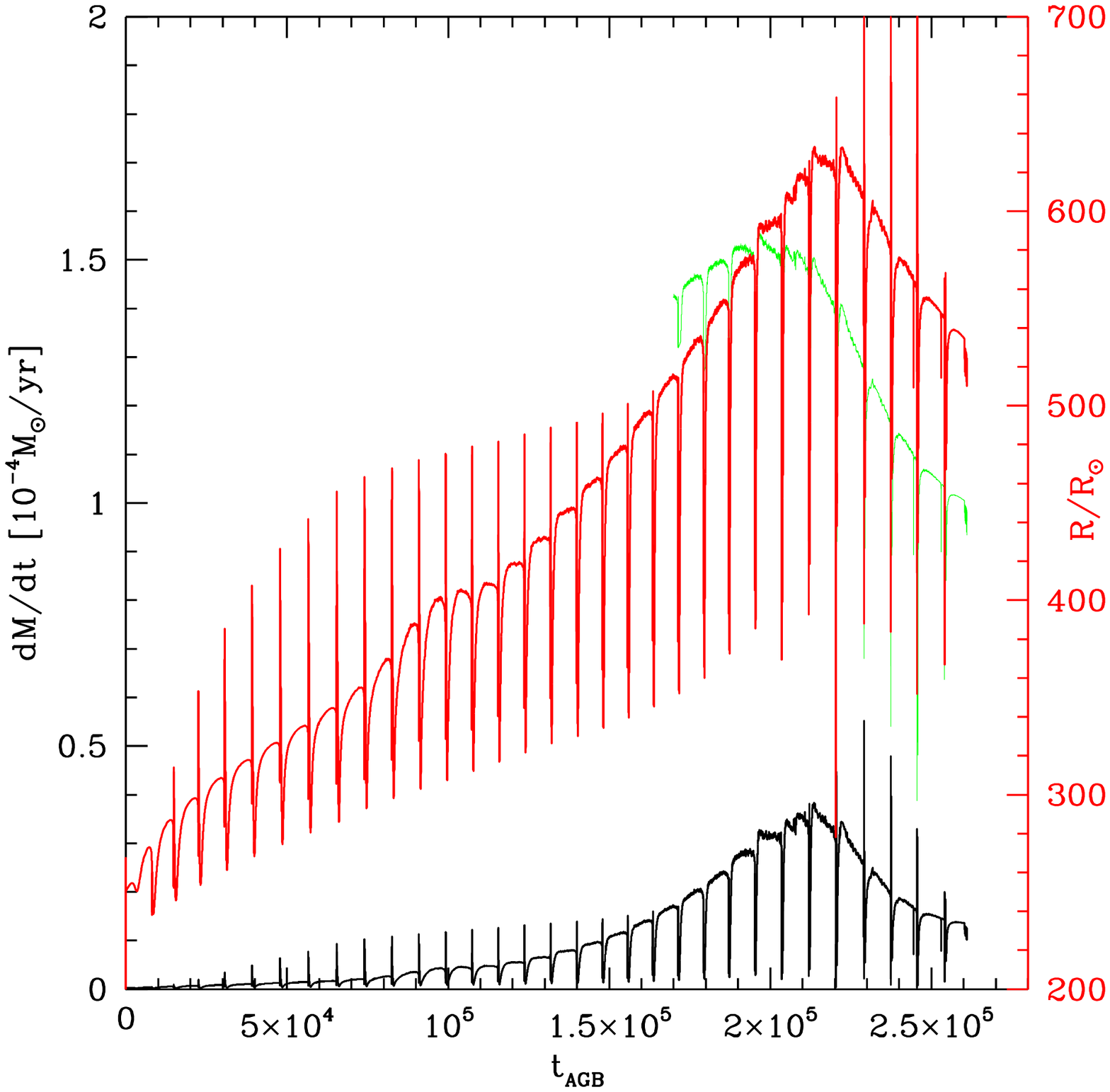}}
\end{minipage}
\vskip-60pt
\caption{The evolution of a $4~{\rm M}_{\odot}$ model star of metallicity $Z=4\times 10^{-3}$,
in terms of the time variation of some physical quantities. Mass loss was modelled 
according to \citet{blocker91}. Times on the x-axis
are zeroed at the beginning of the AGB phase. The left panel shows the variation of
luminosity (black line, scale left y-axis) and pulsation period (red, scale on the right). 
The right panel shows the evolution of the mass loss rate (black line) and 
of the stellar radius (red, scale on the right). The green line on the right panel
shows the mass-loss rate that would be obtained during the super wind phase if the \citet{vw93} 
prescription for mass loss was adopted.
}
\label{fid27}
\end{figure*}

The stars experiencing HBB descend from ${\rm M} \geq 4~{\rm M}_{\odot}$ progenitors,
which evolve on core masses above $\sim 0.8~{\rm M}_{\odot}$ \citep{ventura13}.
As an example, we show in Fig.~\ref{fid27} the evolution of a $4~{\rm M}_{\odot}$ model star.
These stars are characterized by a typical time variation of the luminosity (see left
panel of Fig.~\ref{fid27}), which
increases during the first part of the AGB evolution, as the core mass increases,
then diminishes during the final AGB phases, when HBB is gradually
turned off, owing to the consumption of the convective envelope \citep{ventura13, karakas14}.

The mass loss rate of this class of stars is rather uncertain: use of different prescriptions
leads to results that differ significantly in the values
of the mass loss rate experienced and in the way it changes along the AGB evolution. 

Overall, massive AGBs are efficient dust manufactures. The dust formed is mostly under 
the form of silicates, with traces of alumina dust and solid iron 
\citep{ventura12, ventura14, flavia14a}. This holds in the case of solar or sub-solar chemical 
compositions only, because the scarcity of silicon and aluminium prevents significant dust 
formation during the evolution of metal-poor stars. 

When the \citet{blocker95} treatment of mass loss is adopted, given the tight relationship
between $\dot{\rm M}$ and luminosity in the \citet{blocker95} formula, we find that
the largest mass loss rates take place in conjunction with the largest luminosity (compare
the time variation of the luminosity and of the mass loss rate in the left and right panel
of Fig.~\ref{fid27}, respectively);
this will also be the phase when the dust production rate is highest, because the
rate of mass loss affects directly the density of the wind \citep{fg06}, thus the 
number of gaseous molecules available to condense into dust.

To model ID 27 we considered a $4~{\rm M}_{\odot}$ model star of metallicity
$Z=4\times 10^{-3}$, consistently with the metallicity given by \citet{devika22a}.
As shown in the left panel of Fig.~\ref{fid27}, this star reaches a peak luminosity 
$\sim 35000~{\rm L}_{\odot}$, which decreases down to $\sim 22000~{\rm L}_{\odot}$ during the 
final phases, preceding the contraction to the post-AGB evolution. Before the start of the 
contraction phase the model star is characterized by an effective temperature ${\rm T}_{\rm eff} = 3400$ K, 
radius $\sim 500~{\rm R}_{\odot}$, period $\sim 1200$ d and mass $0.87~{\rm M}_{\odot}$. 

In the right panel of Fig.~\ref{fid27} we can see that the largest mass loss rate found when 
the \citet{blocker91} treatment is used is $\dot{\rm M} \sim 5\times 10^{-5}~{\rm M}_{\odot}/$yr, 
whereas during the final AGB phases it decreases to $\dot{\rm M} \sim 10^{-5}~{\rm M}_{\odot}/$yr. 
Dust formation modelling based on this mass loss rate, combined with the values of luminosity, 
effective temperature and mass given above, and with the surface chemistry of the 
star\footnote{The dust formed in the envelope of oxygen-rich stars is determined by the surface iron, 
silicon and aluminium mass fractions; because the abundance of these elements does not change 
during the AGB phase, the initial chemical composition of the star is relevant for the amount of dust 
formed}, leads to a TAGB optical depth $\tau_{10}=0.4$. Using Equation 4 of Paper I, which
connects the nowadays optical depth with the $\tau_{10}$ characterizing the star when 
the dust was lastly released, we find that the current optical depth should be
$\tau_{10} \sim 5\times 10^{-3}$, far lower than the value of $\tau_{10}$
derived for ID 27, reported in Table \ref{tabpost}.

This result confirms recent findings by Marini et al. (to be submitted to A$\&$A), that
use of the mass loss law by \citet{blocker95} during the final AGB phase hardly allows to
interpret the SED of dust enshrouded AGB stars in the Galaxy, likely descending from massive AGBs 
undergoing HBB, now evolving through the late AGB stages. On the other hand, Marini et al. 
found that use of the \citet{vw93} 
treatment allows a much more satisfactory interpretation of the observations of these kind of sources.
The application of the \citet{vw93} recipe, when
the values of luminosity, effective temperature and period given above are considered,
leads to TAGB optical depths $\tau_{10} \sim 5$, which corresponds to
nowadays optical depths consistent with the value found for ID 27 (see Table \ref{tabpost}).

When using these enhanced mass-loss rates during the very final AGB phases,
we find that the time interval between the tip of the AGB and the current epoch
is $\sim 2000$ yr. If we consider a later phase, when the effective temperature is
3500 K, the time interval is $\sim 400$ yr. If we assume a velocity $\sim 20$ Km$/$s
we find that the distance travelled from dust corresponds to the value derived
from SED fitting.

We note that this conclusion is independent of the way mass loss is described during the 
whole AGB evolution. The treatment of mass loss is definitively relevant for the duration
of the whole AGB phase, and affects the number of TP experienced by the star; 
however, the conditions during the very final AGB stages, after HBB is turned off, are 
practically independent of the previous AGB evolution, being mainly connected with the 
value of residual core mass when the final contraction begins. 

The optical depths derived from these sources are consistent with the above
discussion. For ID 27 we find $\tau_{10}=0.7$, whereas for ID 21 we derive
$\tau_{10}=0.3$ from SED fitting. The difference between the two stars is
mostly related to the lower metallicity of ID 21 than ID 27, thus to the lower
quantities of silicon atoms available.

\subsection{Dust-free stars}
The sources ID 22 and 25 investigated by \citet{devika22a} show no 
infrared excess (see Fig.~\ref{fseddustfree}), which indicates that there is no 
dust in their surroundings, thus no dust production occurred during the
final part of the AGB phase. Indeed they populate the lower region of the plane shown in the left panel of
Fig.~\ref{fsumm}, where we also find ID 23 and ID 31 whose SED exhibit small IR excess, corresponding optical depth $\tau_{10}\leq10^{-3}$. These sources are not reported on
the right panel of the figure, as it is not possible to locate the dusty layer.

For what concerns the sources ID 22, 23 and 31, the lack of dust is
due to the low metallicity (see Table~\ref{tabpost}), which prevents the formation of 
significant quantities of silicates and of alumina dust. This conclusion is general, it is  
independent of the progenitor's mass and on the current luminosity, and stems from the
fact that the formation of dust in the winds of oxygen-rich stars, unlike their
carbon-rich counterparts, is extremely sensitive to the metallicity of the star
(see van Loon 2000
for an observational point of view, while the theoretical predictions are discussed in  Ferrarotti \& Gail 2006 and Ventura et al.
2014).

This explanation cannot be applied to ID 25, because the metallicity is
$[$Fe$/$H$]=-0.4$. The luminosity is consistent with $\sim 4~{\rm M}_{\odot}$
progenitor, as discussed in \citet{devika22b} (note that this star is flagged as Q1,
thus the luminosity can be considered as fairly reliable). Stars of similar masses and metallicity 
experience HBB
during the AGB evolution, and are expected to produce significant quantities of silicates in 
their circumstellar envelope ($\sim 2\times 10^{-3}\rm{M}_{\odot}$ during the entire AGB phase, see 
e.g. Ventura et al. 2014), which should be detected nowadays. The signature of proton-capture
nucleosynthesis is confirmed by the large N enhancement, with $[$N$/$Fe$]$ slightly below unity,
and the large sodium detected, $[$Na$/$Fe$]=+0.8$ \citep{luck90}.

One of the results obtained in Paper I is that bright stars are characterized by optical
depths generally higher than their lower luminosity counterparts surrounded by the same
kind of dust, because the evolutionary times, particularly the transition from the AGB to the 
post-AGB phase, are shorter, thus the dust released is closer to the central object. All these 
factors render the post-AGB nature of ID 25 extremely unlikely. It is more likely the 
original classification by \citet{luck90}, also considered by \citet{hans97}, that this star is a supergiant, and that the surface N and Na increased as a consequence of meridional currents active 
during the main sequence phase, triggered by the rotation of the star \citep{limongi18}.

\subsection{Low-luminosity, AGB-manque stars?}
The stars ID 19 and ID 20 are currently evolving at luminosities below the
threshold required to start the TP phase \citep{devika22a}. \citet{devika22b}
suggested that these objects descend from low-mass progenitors, which are
evolving through the post-HB phase, and have started the contraction to the blue,
after evolving to the red. In this framework, it is during the latter phase
that the dust nowadays observed was released, as no meaningful dust formation is 
expected for effective temperatures above 4000 K.

While on the evolutionary point of view this interpretation is consistent
with results from stellar evolution modelling, the infrared excess exhibited by
ID 19 and ID 20, which corresponds to $\tau_{10}=0.2$ (see Fig.~\ref{fsumm}), 
is significantly higher than expected. Indeed the mass-loss rates obtained from
stellar evolution modelling are slightly above 
$10^{-7}~{\rm M}_{\odot}/$yr, which corresponds to optical depth below 0.05.
The application of equation 4 of Paper I allows to determine the current optical depth, 
$\tau_{10} \sim 0.01$, which is inconsistent with the values derived from SED fitting.

ID 19 is a Q1 star, with a derived luminosity far below the threshold required to
reach the TP phase. This suggests that it completed core helium burning, then evolved 
as a giant, before the envelope, of a few $0.01~{\rm M}_{\odot}$, was lost, and the 
evolutionary track moved to the blue side of the HR diagram. The contraction process 
must have been rather quick, owing to the little mass of the envelope, until the present 
day effective temperature is reached. This is consistent with the results shown in the 
right panel of Fig.2, where we notice that ID 19 is the star with the nearest dusty layer, 
below $10^4$ solar radii.

ID 20 was flagged as Q2 by \citet{devika22a}, which leaves some room for invoking
a higher luminosity for this source. We believe possible that the luminosity is
underestimated, and that the true value is something around $4000~{\rm L}_{\odot}$.
If thus hypothesis proves correct, the star must have entered the AGB when the envelope 
mass was below $0.1\rm M_{\odot}$, and then it was lost rapidly, likely
already during the first thermal pulse. This is on the wake of the scenario first
proposed by \citet{renzini89}, which relies on the relatively long duration of the
peak luminosity experienced by very low-mass stars. 

An alternative explanation is based on the fact that the nitrogen abundance of ID 20
given in \citet{devika22a} is consistent with the results from HBB nucleosynthesis, 
when the upper values are considered. In this case ID 20 would descend from a massive AGB
progenitor, which experienced HBB. Only a better determination of the parallax
will allow a better understanding of this object.

\subsection{A few outliers}
The sample of oxygen-rich stars with no s-process enrichment by \citet{devika22a}
includes three sources for which a full understanding of the evolutionary and dusty
properties can be barely reached.

ID 26 is a peculiar object, as it is significantly enriched in nitrogen, despite the low
luminosity. We note that this source is flagged as Q1 by \citet{devika22a}, thus
the luminosity range given is reliable. \citet{devika22b} suggested that ID 26
experienced deep mixing during the RGB phase, which might explain the nitrogen
and sodium enrichment measured.

On the other hand this source is characterized by a large IR excess, and the
derived optical depth is $\tau_{10}=0.3$, among the highest found in the present
analysis. The occurrence of extra-mixing is not expected to affect the dust
production during the final phases, which is relevant to determine the dust
surrounding the star during the post-AGB. The dusty properties of this source
suggest that the star descends from progenitors which experienced HBB.
However, this result must be taken with some caution, given the claimed
asymmetry in the dust distribution \citep{min2013}, which might clash
with the plain interpretation obtained here, based on the adoption of the 
DUSTY code in the isotropic modality.

In summary, while the surface chemistry and the dust properties of ID 26
are consistent with progenitors of mass above $\sim 3~{\rm M}_{\odot}$, which
experienced HBB and form significant amount of dust until the end of the
AGB phase, the luminosity is compatible with a low-mass progenitor.

The optical depth and the distance of the dusty layer from the star of ID 28 suggest that this source descends from an intermediate mass progenitor. This interpretation could work if the upper limit for the metallicity reported in Tab. 1 is considered. We insert this source among the outliers because the only derivation of the surface nitrogen by \citet{sahin11} appears to be inconsistent with the fact that ID 28 should have experienced HBB, which would lift the surface N by $\sim$ 1 dex.

ID 29 is one of the objects characterized by large optical depth, 
$\tau_{10}=0.7$. Unfortunately the luminosity of this source, flagged 
as Q2, is highly uncertain, with 
$4300~{\rm L}_{\odot} < {\rm L} < 22800~{\rm L}_{\odot}$ \citep{devika22a}. Fruthermore the surface abundances of nitrogen and oxygen are not available. This prevents any identification of the progenitor and the interpretation of the IR excess. 
A further criticality related to this source is the possible presence
of a disk, which would render the methodology used in the present analysis
unapplicabile. \citet{pereira04}, based on the morphology of the SED, suggested
that the star is surrounded by a dust disk. This conclusion is consistent
with the results obtained by \citet{goldman2017}. Indeed, ID 29 appears to be the only
source whose SED does not show the double-peak structure (see Fig.~\ref{fsed}), typical of 
spherically symmetric sources. This is the reason for the poor agreement between the fit and the data, in particular in the region of the spectrum below $\lambda\sim11\mu m$, regardless the choice on the percentage of crystalline silicate considered.

\section{Discussion}
\label{disc}
The infrared excess characterizing post-AGB stars is related to the presence of dust
in the surroundings of the stars, which was released during earlier epochs. On general
grounds, the optical depth of the dusty region is determined by the rate of dust formation
and by the current distance of the dusty region from the surface of the star; this
distance is determined by the evolutionary time scale of the AGB - to post-AGB phase transition,
and by the velocity with which the outflow moved away from the star after the dust was
lastly released.

The situation for oxygen-rich stars emerging from this study is complex, as can be understood by looking
at Fig.~\ref{fsumm}, where it is clear that in this case luminosity, unlike the carbon stars investigated 
in Paper I, is not the only factor affecting the dust properties of post-AGB stars. 
Metal-poor oxygen-rich stars form little or no dust, independently of luminosity. 
If we restrict the attention on the sources with solar or sub-solar chemical composition,
a rough correlation between luminosity and optical depth can be seen in the left panel of
Fig.~\ref{fsumm}: the stars distribute on this plane along a diagonal band, covering the luminosity 
range $4000-18000~{\rm L}_{\odot}$ and the optical depths $5\times 10^{-3} < \tau_{10} < 0.9$.
ID 19 is off this trend, which might be connected to the fact that this star did not
evolve through the AGB phase.

The optical depth of stars descending from 
massive AGBs that experienced HBB is generally higher than the low-mass counterparts of
similar metallicity, owing to the larger rates of dust production experienced during the
whole AGB lifetime, and particularly during the final AGB phases. However, inspection
of the right panel of Fig.~\ref{fsumm} shows that the relationship between the distance of the
dusty layer from the surface of the star and the luminosity is less obvious than we found 
in Paper I for carbon stars in the MC.

We believe that the main reason for this is the intrinsic difference between the extinction
properties of carbon dust and silicates, the mean opacities of carbonaceous dust being
one order of magnitude higher than silicates, under the thermodynamic conditions when dust
is expected to form (see Fig.~10 in Ferrarotti \& Gail 2002). The scattering and absorption coefficients of
carbon dust are extremely large, thus the formation of carbon dust, even in moderate
quantities, favours a significant acceleration of the outflow: in these conditions, the
current location of the dusty region is mainly determined by the time scale of the transition
from the AGB to the post-AGB phase, thus by the luminosity of the star.

In the case of the formation of silicates, it is not obvious that radiation pressure is able to overcome the
gravitational pull of the central star, and to accelerate the wind: this reflects into a higher
heterogeneity in the outflow velocities of oxygen-rich stars with respect to carbon stars. 
As far as bright O-rich stars which experienced HBB are considered, the luminosities, around or above 
$2\times 10^4~{\rm L}_{\odot}$, are sufficiently large that the radiation pressure is
strong enough to drive the wind. This is not as much obvious in the low-luminosity
($L < 6000~{\rm L}_{\odot}$) domain of M-type stars.
Indeed in the 
low-luminosity region of the plane shown in the right panel of Fig.~\ref{fsumm}, populated by 
the progeny of low-mass stars that failed to reach the C-star stage, we note that the 
distance of the dusty region from the star is in fact correlated with the luminosity: the largest 
distance is found for ID 30, the brightest among these objects. A possible explanation for this 
behaviour is that by the combined analysis of the 5 oxygen-rich stars discussed in Paper I and
of ID 24 and ID 30 we are exploring the transition from winds scarcely affected by radiation
pressure, to radiation driven winds. This trend reflects into
a positive correlation between velocity and luminosity, which is the reason for the larger
distances of the dusty regions in brighter stars.

We note in the right panel of Fig.~\ref{fsumm} that the distance of the dusty regions of
ID 30 is similar to those of the brighter, higher mass counterparts ID 21 and ID 27. 
We believe that even in this case the 
explanation can be found in the different expansion velocities of the outflows, which, in
turn, are connected to the large differences between the dust production rates by low-mass
and massive AGBs.
In the case of the carbon stars investigated in 
Paper I, the difference in the mass loss rates during the final AGB phases between the brighter stars
and the low-luminosity C-stars was a factor $\sim 3$. In this case we find that almost two orders of magnitude
separate the mass loss rate experienced by ID 27 ($\sim 10^{-4}{\rm M}_{\odot}/$yr)
and ID 30 ($\sim 2\times 10^{-6}{\rm M}_{\odot}/$yr). Since for mass conservation reasons
the velocity of the outflow is tightly connected with the mass-loss rate, 
the outflow velocity is significantly larger than in low-mass
counterparts, and this effect counterbalances the differences in the evolutionary time scales,
on the determination of the location of the dusty region.

\section{Conclusions}
\label{concl}
We study the properties of the dust surrounding oxygen-rich post-AGB stars, to reconstruct the dust 
formation process during the late phases of the AGB evolution, which prove the most relevant to understand
the dust budget expected from low and intermediate mass stars. To this aim, we focus on oxygen-rich
post-AGB sources observed in the Galaxy \citep{devika22a}, whose distances (and hence luminosities) can be inferred from the Gaia data release, 
for which the SED can be reconstructed with sufficient accuracy by the combined availability of results 
from optical and infrared photometry.

To reconstruct the evolutionary and dust formation history of these stars, we follow an approach based
on the derivation of the main dust and stellar parameters via SED fitting, and on the interpretation of
the results on the basis of stellar evolution and dust formation modelling of the AGB and post-AGB
phases. This allowed us to identify: a) sources descending from low-mass progenitors, which loose the
external mantle before experiencing a number of TDU episodes sufficient to rise the surface s-process
content and to reach the C-star stage; b) the progeny of massive AGB stars that experienced HBB, which
is further witnessed by the nitrogen enhancement derived from high resolution spectroscopy analysis;
c) stars with little or no IR excess, which we connect to the scarcity of dust in their
surroundings; d) faint sources, likely evolving through the post-HB phase, which failed
to reach the AGB phase, owing to small mass of the envelope at the end of the core
helium burning.

If we restrict our attention only to the stars belonging to the groups (a) and (b)
mentioned above, we find that those of higher luminosity are surrounded by a more optically
thick dusty region, which is consistent with results from stellar evolution and dust formation 
modelling, according to which brighter stars of a given metallicity experience higher mass loss
and dust production rates during the final AGB phases. Indeed in the case of massive AGBs such
a comparison between the observational evidence and the results from modelling allow to draw 
information on the still largely unknown mass loss mechanism experienced by these stars when evolving
through the AGB phase, indicating that the rate of mass loss keeps above $\sim 10^{-4}~{\rm M}_{\odot}/$yr
until the beginning of the transition from the AGB to the post-AGB phase.

An exception to this general behaviour is
provided by a subset of metal poor stars, which have little or no dust in their surroundings,
owing to the scarcity of silicon and aluminium in the surface regions. A further exception to this
rule is the bright source ID 25, with sub-solar chemical composition and no dust, which we suggest to be
a red supergiant star.

The analysis of the distribution of the distances of the dusty regions from the central stars, 
derived from SED fitting, offers an interesting opportunity to derive information regarding the
dynamics of the winds of oxygen-rich stars, during the transition from the AGB to the post-AGB phase.

In the low-mass domain the distance of the dusty region is correlated to the luminosity, 
since brighter stars experience higher rates of dust formation than the lower luminosity
counterparts of similar metallicity, thus the outflow moves faster, owing to the higher
radiation pressure acting on the dust grains in the circumstellar envelope.
The analysis of the stars descending from massive AGB progenitors shows that the
dusty regions are located at larger distances with respect to the ones found for the brightest low-mass stars: this is
explained by the balance between the effects of the shorter transition time scales of 
higher luminosity stars and the larger dust production, which reflects into faster
outflows.

\label{end}

\begin{acknowledgements}
We thank the referee Dr. Jacco Th. Van Loon for his prompt, detailed and constructive revision of the manuscript. DK  acknowledges the support of the  Australian Research Council (ARC) Discovery Early Career Research Award (DECRA) grant (DE190100813). This research was supported in part by the Australian Research Council Centre of Excellence for All Sky Astrophysics in 3 Dimensions (ASTRO 3D), through project number CE170100013. HVW acknowledges support from the Research Council of the KU Leuven under grant number C14/17/082. EM acknowledges support from the INAF research project “LBT - Supporto Arizona Italia".

\end{acknowledgements}

%
%

\end{document}